\documentclass[aps,prl,showpacs,twocolumn]{revtex4-2}%
\usepackage{graphicx}
\usepackage{graphicx}
\usepackage{amsmath}
\usepackage{amsfonts}
\usepackage{xcolor}
\usepackage[T2A]{fontenc}
\usepackage{amssymb}
\newcommand{\pd}[3][]{\frac{\partial{#2}}{\partial{#3}}}
\newcommand{\pdd}[3][]{\frac{\partial^2{#2}}{\partial{#3^2}}}
\providecommand{\LyX}{L\kern-.1667em\lower.25em\hbox{Y}\kern-.125emX\@}

\begin{document}
\title{From Delay to Inertia and Triadic Interactions: {A Predictive Model for\\ Time-Delayed Oscillator Networks}}
%\title{From Delay to Inertia and Triadic Interactions: A State-predictive for Time-Delayed Oscillator Networks}

\author{Lev A. Smirnov$^{1}$, Vyacheslav O. Munyayev$^{1}$, Maxim I. Bolotov$^{1}$, and Igor Belykh$^{2}$\footnote{Corresponding author, e-mail: ibelykh@gsu.edu}}

\affiliation{$^1$Department of Control Theory, Lobachevsky State University of Nizhny Novgorod,
	23 Gagarin Avenue, Nizhny Novgorod, 603022, Russia\\
	$^2$Department of Mathematics and Statistics and Neuroscience Institute, Georgia State University, P.O. Box 4110, Atlanta, Georgia, 30302-410, USA}	

%\begin{abstract}
%Time-delayed phase oscillator networks model biological and physical systems with finite signal speeds. Yet existing first-order reductions of time-delayed cooperative dynamics, turning the time delay into phase lags, are primarily restricted to synchronization or to special intrinsic-frequency distributions. This Letter develops a second-order reduction for Kuramoto–Daido oscillators with time-delayed coupling, transforming the original time-delayed system of one-dimensional phase oscillators into a delay-free network of two-dimensional rotators. The resulting model shows that coupling delay generates inertial terms in the intrinsic dynamics and higher-order (triadic) interactions and accurately predicts complex collective patterns such as splay, cyclops, and chimera states. It further reveals that time delay acts primarily as an effective inertia for higher-dimensional dynamics, including splay states. In contrast, the induced triadic interactions are decisive in lower-dimensional patterns, such as chimera states. The method applies to networks with arbitrary topology, higher-harmonic coupling, and intrinsic-frequency heterogeneity, yielding a compact, parameter-explicit reduced model. This universal description of time-delayed oscillator networks opens the door to systematic prediction and analysis of nontrivial collective dynamics in delay-coupled systems.
%\end{abstract}
\begin{abstract}
	Time-delayed {oscillator} networks {underlie} diverse biological and physical systems, yet standard first-order phase reductions {fail to capture } their high-dimensional collective dynamics. In this Letter, we develop a {universal second-order predictive reduction} for {time-delayed} Kuramoto--Daido networks that {maps delayed one-dimensional phase dynamics to} a delay-free network of two-dimensional rotators. {Delay induces effective inertia and triadic interactions, yielding accurate predictions of nontrivial attractors and their collective-state statistics, including splay, cyclops, and chimera states.} The reduction {reveals a division of roles: inertia organizes higher-dimensional dynamics, whereas triadic terms are crucial for lower-dimensional patterns such as chimeras.} {Applicable to arbitrary topology, higher harmonics, and intrinsic-frequency heterogeneity, it provides a compact, parameter-explicit reduced model. The same framework also extends to time-delayed amplitude-phase oscillator networks, including swarmalators, yielding analogous reduced equations with emergent inertia and triadic higher-order couplings.} This {unified and readily deployable} description enables systematic prediction and analysis of {delay-controlled} collective dynamics across oscillator networks.
\end{abstract}


\date{\today}
\maketitle

\textit{Introduction.} 
Networks of phase oscillators with time-delayed interactions provide a fundamental framework for collective dynamics in biological, physical, and technological systems with finite signal propagation speeds. A broad literature has shown that such delays can induce rich cooperative behavior, ranging from synchronization and bistability \cite{izhikevich1998phase,yeung1999time,earl2003synchronization,jorg2014synchronization,restrepo2019competitive,ocampo2024strong,bick2024time,lee2009large} to phase and amplitude chimeras, glassy phase-locked states and clustered chimera patterns \cite{bick2017robust,ameli2021time,gjurchinovski2017control,sheeba2009globally,lee2011dynamics}, as well as twisted waves and related structured states \cite{laing2016travelling,an2024stability}, delay-tunable multistability, and topology-dependent transitions \cite{sawicki2017chimera,wu2018dynamics} in Kuramoto, phase-amplitude oscillator and spatially extended networks \cite{wolfrum2006eckhaus,sethia2008clustered,ares2012collective,yanchuk2014pattern}. Comparable delay-induced phenomena have been documented in neuronal systems \cite{dhamala2004enhancement,sun2017effects,lehnert2011loss,popovych2011delay,keane2012synchronisation,sawicki2019delay,sawicki2019chimeras,lucchetti2021emergence,smirnov2024synaptic} and laser networks \cite{kozyreff2000global,topfer2020time,heiligenthal2011strong}, where heterogeneous time delays can even generate robust disorder-induced phase locking of laser oscillators \cite{nair2021using,barioni2025interpretable}.

Existing first-order reductions of time-delayed systems typically replace the delay by a static phase lag. While effective for synchronization and simple phase-locking \cite{izhikevich1998phase,yeung1999time,earl2003synchronization}, such reductions cannot adequately capture nontrivial collective states, such as splay, clustered, and chimera states, as well as their stability boundaries and basins of attraction. In the thermodynamic limit, the Ott–Antonsen ansatz \cite{ott2008low} yields low-dimensional macroscopic reductions \cite{lee2009large,lee2011dynamics,laing2016travelling} which allow rigorous stability analysis, including delay-induced twisted states, but are restricted to purely first-harmonic coupling and specific choices of frequency and delay distributions. {Higher-order reductions have been developed for non-delayed \cite{rosenblum2019numerical,mau2023high,gengel2020high,mau2024phase} and delayed oscillator pairs \cite{ocampo2024strong}, delay-coupled Stuart--Landau oscillators \cite{bick2024time}, and phase-amplitude descriptions of perturbed delayed limit cycles \cite{Kotani2020,Nicks2024}. However, these approaches are formulated for low-dimensional or oscillator-level systems and rely on system-specific conjugacy equations, making them unsuitable for predicting high-dimensional collective states, statistics, or attraction basins in large networks.} Thus, despite significant progress, we still lack a general predictive {network-level} reduction for time-delayed phase oscillators that accommodates arbitrary topology, heterogeneous frequencies, and multi-harmonic coupling while remaining accurate for high-dimensional dynamics {\footnote{After submission of the present manuscript, a related preprint by N.~Fujii, K.~Taga, R.~Muolo, B.~Rink, and H.~Nakao (arXiv:2512.16193) appeared deriving a phase-only effective model with two-body and three-body interactions for globally coupled Kuramoto oscillators, with emphasis on synchronization transitions and Ott--Antonsen analysis. In the special case of sinusoidal coupling and weak frequency heterogeneity, Eq.~(S.12) of our Supplementary Material is closely related in structure to their reduced equations. In contrast, the present work develops a compact inertial second-order reduction for broad time-delayed Kuramoto--Daido networks with arbitrary topology, heterogeneous frequencies, and higher-harmonic coupling, aimed at predicting high-dimensional collective states and their statistics.}}.

In this Letter, we close this gap and develop a second-order reduction for a broad class of time-delayed Kuramoto–Daido (KD) networks \cite{acebron}, transforming the original delayed system of one-dimensional (1D) phases into a delay-free network of two-dimensional rotators. The reduced model makes the effects of delay explicit through inertially augmented intrinsic dynamics and triadic phase interactions. It quantitatively reproduces complex collective states such as splay and chimera patterns in KD networks with time-delayed global and Kuramoto–Battogtokh–type nonlocal coupling \cite{smirnov2017chimera}. We further show that the reduced model enables a constructive analysis of cyclops states \cite{munyayev2023cyclops}—a distinct class of three-cluster generalized splay states—in networks with higher-harmonic coupling. Remarkably, the robust emergence of these cyclops states relies on genuinely two-dimensional intrinsic oscillator dynamics generated by delay-induced inertia, and is therefore inaccessible to standard first-order 1D phase reductions.

\textit{Time-delayed model and its second-order reduction.} 
We study a generalized KD network with heterogeneous natural frequencies, external forcing, and time-delayed pairwise coupling:
\vspace{-3mm}
\begin{equation}
	\dfrac{d\theta_{j}(t)}{dt} = \varpi + \eta_{j}(t )+ \frac{\varkappa}{N}\sum_{k=1}^{N}F_{jk}\bigl(\theta_{k}(t-\tau)-\theta_{j}(t)\bigr),
	\label{eq:kd-model}
	\vspace{-3mm}
\end{equation}
where $\theta_j(t)$ is the phase of oscillator $j$ ($j=1,\dots,N$), $\varpi$ is a baseline (mean) natural frequency, $\eta_j(t)$ collects deviations from this baseline (including external forcing or detuning), $\kappa$ is the coupling strength, and $F_{jk}(\cdot)$ are $2\pi$-periodic pairwise coupling functions that may differ across oscillator pairs{, for example, through single-harmonic, higher-harmonic, or more general periodic interactions. 
	%In particular, random or heterogeneous network structure can be incorporated through link-dependent couplings, e.g., $F_{jk}(\vartheta)=A_{jk}H_{jk}(\vartheta)$, where $A_{jk}$ is a weighted or random connectivity matrix}. 
The parameter $\tau>0$ is a uniform coupling delay. The results extend straightforwardly to heterogeneous node-wise delays $\tau_j$, provided all incoming connections to node $j$ share the same delay $\tau_j$.

We analyze system (1) in the regime of weak frequency heterogeneity and coupling by introducing a small parameter $\varepsilon \ll 1$ so that $\eta_j(t)=\varepsilon\omega_j+\varepsilon\zeta_j(t)$, $\varkappa=\varepsilon\kappa$, where $\omega_j$ are small detunings from the baseline frequency $\varpi$ and $\zeta_j(t)$ is a zero-mean time-dependent perturbation, {deterministic or stochastic, e.g., a temporally correlated (colored) random forcing}.

In the limit $\varepsilon=0$, all oscillators rotate uniformly with frequency $\varpi$. Treating $\varepsilon$ as a small parameter, we perform a lengthy multiple–time-scale expansion to capture the slow modulation of phases by weak disorder and coupling (see the Supplementary Material for details). We introduce slow times $t_s=\varepsilon^s t$ ($s=0,1,2,\ldots$) and write
\vspace{-3mm}
\begin{equation}
	\theta_j(t)\!=\!\varpi t_0 + \phi_j(t_1,t_2,\dots) +
	\sum_{p=1}^\infty \varepsilon^p
	\varphi_j^{(p)}(t_0,t_1,t_2,\dots),
	\label{eq:series}
	\vspace{-3mm}
\end{equation}
where $\varpi{t_{0}}$ represents the leading-order, fast oscillatory behavior, $\phi_j$ describes the slow phase dynamics, and $\varphi_j^{(p)}$ are higher-order fast corrections that average out over the longer-term evolution of the system.
Substituting \eqref{eq:series} into \eqref{eq:kd-model}, expanding the delayed terms, and eliminating secular contributions yields a solvability hierarchy in powers of $\varepsilon$. At $O(\varepsilon)$, we obtain the familiar delay-free KD phase model with a delay-induced phase shift \cite{izhikevich1998phase}
\vspace{-3mm}
\begin{equation}
	\frac{d\phi_{j}}{dt}=\bar{\eta}_{j}(t)+\frac{\varkappa}{N}
	\sum \limits_{k=1}^{N}F_{jk}\bigl(\phi_{k}-\phi_{j}-\varpi\tau\bigr),
	\label{eq:reduced-model-1}
	\vspace{-3mm}
\end{equation}
where $\bar{\eta}_{j}(t)=\varepsilon\omega_{j}+\varepsilon\bar{\zeta}_{j}(t)$ is the averaged frequency perturbation. At $O(\varepsilon^2)$, the delay couples the slow time derivatives, and in the original time variable, the combined phase dynamics take the form of a second-order equation
\vspace{-2mm}
\begin{multline}
	%\begin{array}{lcl}
	\tau \frac{d^2\phi_{j}}{dt^2}+\frac{d\phi_{j}}{dt}=
	\left[\bar{\eta}_{j}(t)+\frac{\varkappa}{N}\sum \limits_{k=1}^{N}
	F_{jk}\bigl(\mathit{\Delta}_{jk}\bigr)\right]\times\\
	\times\left[1-\frac{\tau \varkappa}{N}\sum \limits_{k=1}^{N}
	F'_{jk}\bigl(\mathit{\Delta}_{jk}\bigr)\right]+
	\tau \frac{d\bar{\eta}_{j}}{dt},
	%\end{array}
	\vspace{-5mm}
	\label{eq:reduced-model-2}
\end{multline}
where $\mathit{\Delta}_{jk}=\phi_{k}-\phi_{j}-\varpi\tau.$ The reduced model \eqref{eq:reduced-model-2} shows that the time delay $\tau$ manifests as an effective inertia, transforming the originally overdamped phase dynamics with time delay into a second-order system. The delay also generates higher-order (triadic) interactions: in Eq.~\eqref{eq:reduced-model-2}, the second-order correction appears as a product of two sums, which yields a double sum corresponding to triplet couplings. {The reduction also remains valid in the presence of weak external or random perturbations, which enter explicitly through $\bar{\eta}_j(t)$ and, at second order, through $\tau\, d\bar{\eta}_j/dt$.} {The interplay of delay-induced inertia and higher-order interactions} becomes particularly transparent for identical oscillators, where we set $\bar{\eta}_{j}(t)=0$ (assumed hereafter) and consider single-harmonic coupling $F_{jk}(\vartheta)=G_{jk}\sin(\vartheta-\alpha)$ and $F'_{jk}(\vartheta)=G_{jk}\cos(\vartheta-\alpha)$,
with $G_{jk}$ denoting the coupling weights.
Using trigonometric identities and the cancellation of antisymmetric double sums,
Eq.~\eqref{eq:reduced-model-2} can be rewritten as 
\vspace{-4mm}
\begin{multline}
	%\begin{array}{l}
	\tau \frac{d^2\phi_{j}}{dt^2}+\frac{d\phi_{j}}{dt}=
	\frac{\varkappa}{N}\sum \limits_{k=1}^{N}G_{jk}\sin\left(\phi_{k}-\phi_{j}-\tilde{\alpha}\right)-\\
	-\frac{\tau\varkappa^{2}}{2N^{2}}\sum \limits_{k=1}^{N}\sum \limits_{\ell=1}^{N}
	G_{jk}G_{j\ell}\sin(\phi_{k}+\phi_{\ell}-2\phi_{j}-2\tilde{\alpha}),
	\label{eq:reduced-model-2b}
	\vspace{-5mm}
	%\end{array}
\end{multline}
where $\tilde{\alpha}=\alpha+\varpi\tau$. The first sum corresponds to the standard pairwise Kuramoto–Sakaguchi coupling with a delay-induced phase shift. In contrast, the second sum is a purely delay-induced triadic interaction that couples oscillator $j$ to phase triplets $(j,k,\ell)$. This structure mirrors the triplet interactions obtained via higher-order phase reduction for non-delayed Stuart–Landau oscillators \cite{mau2024phase}, but here arises solely from memory (delay) without introducing amplitude degrees of freedom.

{The reduced model \eqref{eq:reduced-model-2} gives a delay-free inertial representation of the delayed network. We now compare it with the original delayed system and the first-order reduction across global, nonlocal, small-world, and biharmonic couplings, focusing on collective-state statistics, phase profiles, and basins of attraction.}

\textit{Global, one-harmonic coupling.} We first test the reduced model \eqref{eq:reduced-model-2b} for single-harmonic mean-field coupling, $F_{jk}(\vartheta)=G_{jk}\sin(\vartheta-\alpha)$ with $G_{jk}=1$ for all $j,k$. Introducing the Kuramoto order parameters $Z_m=\sum_{k=1}^{N} e^{i m \phi_k}\big/N$, Eq.~\eqref{eq:reduced-model-2b} can be written in the compact mean-field form
\vspace{-1mm}
\begin{equation}
	\tau \frac{d^2\phi_{j}}{dt^2}+\frac{d\phi_{j}}{dt}
	=\varkappa\,\mathrm{Im}\bigl[Z_{1}e^{-i\psi_{j}}\bigr]
	-\frac{\tau\varkappa^{2}}{2}\mathrm{Im}\bigl[Z_{1}^{2}e^{-2i \psi_{j}}\bigr],
	\label{eq:reduced-model-2c}
	\vspace{-1mm}
\end{equation}
where $\psi_j=\phi_{j}+\tilde{\alpha}$ and 
$\tilde{\alpha}=\alpha+\varpi\tau$. Thus, delay induces an effective state-dependent second harmonic proportional to $Z_{1}^{2}$. For high-dimensional collective dynamics with small first-order parameter $r_1=|Z_1|\ll1$ (e.g., generalized splay states \cite{berner2021generalized}), the triadic term is negligible, and the delay acts primarily as effective inertia. In contrast, for lower-dimensional patterns such as chimera states with intermediate $r_1$, the triadic term becomes significant and encodes delay-induced higher-order correlations. The model \eqref{eq:reduced-model-2c} is analytically tractable. For generalized splay states $\phi_j=\phi_j^0=\text{const}$ with $Z_1=\sum_{k=1}^N e^{i\phi_k^0}\big/N=0$, Eq.~\eqref{eq:reduced-model-2c} reduces to the Kuramoto model with inertia \cite{berner2021generalized,munyayev2022stability}. The generalized splay-state stability condition derived in \cite{munyayev2023cyclops} can be written in terms of the parameters of \eqref{eq:reduced-model-2c}
as
%\begin{equation}
$\textcolor{black}{\tilde{\varkappa}<\tilde{\varkappa}^*}={2\cos\tilde{\alpha}}/{(r_2^2-\sin^2\tilde{\alpha})}, \quad \text{for } |\sin\tilde{\alpha}|>r_2$, \textcolor{black}{or $\tilde{\varkappa}>0, \quad \text{for } \cos\tilde{\alpha}<-\sqrt{1-r_2^2}$,} 
%\label{eq:sps_rdc_bnd}
%\end{equation}
where $\tilde{\varkappa}=\tau\varkappa$ and $r_2=|Z_2|$ is the magnitude of the second Kuramoto moment. Similarly, linearizing \eqref{eq:reduced-model-2c} about the fully synchronous state $\phi_j=\varpi t$ yields  the stability condition: \textcolor{black}{ 
	$\tilde{\varkappa}>\tilde{\varkappa}^*=\cos\tilde{\alpha}/\cos(2\tilde{\alpha}), \quad \text{for } |\sin\tilde{\alpha}|>1/\sqrt{2}$, or $\tilde{\varkappa}<\tilde{\varkappa}^*=\cos\tilde{\alpha}/\cos(2\tilde{\alpha}), \quad \text{for } \cos\tilde{\alpha}>1/\sqrt{2}$,} in agreement with classical necessary and sufficient condition \cite{earl2003delay}: 
$\varkappa F'(-\varpi_s\tau)>0$ 
obtained directly from the time-delayed system, up to terms of order $\tilde{\varkappa}^2$.
Figure~\ref{fig:fig1}(a,b) tests the predictive power of the second-order model \eqref{eq:reduced-model-2c} by comparing it with the original delayed system \eqref{eq:kd-model} and the first-order reduction \eqref{eq:reduced-model-1}. Figure~\ref{fig:fig1}(a) shows that the second-order model reproduces the statistics of the delayed system almost perfectly, whereas the first-order model misestimates these probabilities and frequently misses the synchronous state altogether.
Figure~\ref{fig:fig1}(b) compares probability density functions (PDFs) of generalized splay states at fixed $r_2$: the PDFs (and the individual phase configurations, inset) from the second-order and delayed models coincide, while the first-order reduction exhibits a pronounced bias. Figure~\ref{fig:fig1}(c) presents the stability diagram of synchrony and generalized splay states in terms of $\tilde{\varkappa}=\tau\varkappa$ and $\tilde{\alpha}=\alpha+\varpi\tau$. For generalized splay states with $r_2=0$, the numerical stability curves $\Gamma_3$ (the delayed system) and $\Gamma_4$ (second-order system)
begin to diverge around $\tilde{\varkappa}\approx0.5$, indicating the range of quantitative validity of the second-order approximation. Since the effective expansion parameter is $\tilde{\varkappa}=\tau\varkappa$, small delays permit relatively strong coupling, while weak coupling allows predictive use of the reduced model even for comparatively large delays.
\begin{figure}
	\includegraphics[width=0.8\columnwidth]{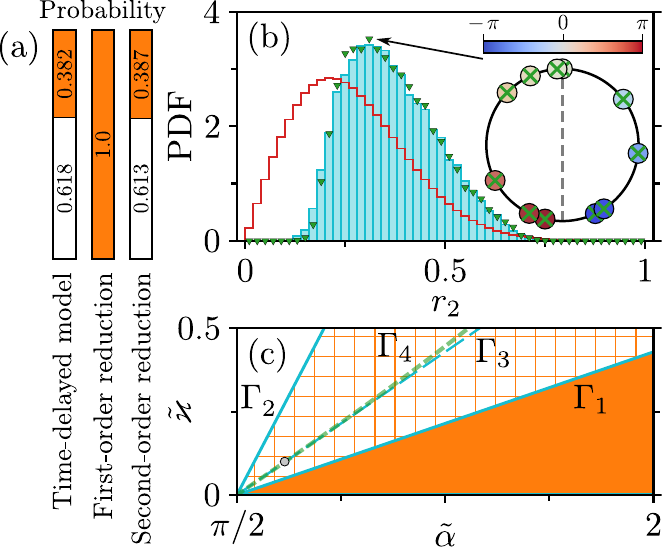}
	\vspace{-2mm}
	\caption{(a) Probabilities of convergence to full synchrony (white) and generalized splay states (orange) for the time-delayed model with global, single-harmonic coupling and its first-order \eqref{eq:reduced-model-1} and second-order \eqref{eq:reduced-model-2c} reductions, estimated from $5\times10^4$ simulations with random initial conditions (uniform phases; for the delayed system, random phase histories on $[-\tau,0]$). (b) Probability density functions of $r_2=|Z_2|$ for realized splay states in the time-delayed system (cyan histogram), compared with the first-order reduction (red curve) and the second-order reduction (green triangles). Parameters in (a,b): $N=11$, $\varkappa=0.1$, $\tau=1$, $\varpi=1$, $\alpha=0.62$ (gray dot in (c)). (c) Stability regions of full synchrony (white) and splay states with $r_2\le0.8$ (orange) for the time-delayed model in the $(\tilde{\varkappa},\tilde{\alpha})$ plane, where $\tilde{\varkappa}=\tau\varkappa$ and $\tilde{\alpha}=\alpha+\varpi\tau$. The hatched region indicates multistability of synchrony and generalized splay states. 
		The solid cyan curves $\Gamma_1$ and $\Gamma_2$ are the stability boundaries for synchrony and splay states in the delayed system (the analytic curve $\Gamma_1$ corresponds to the condition 
		%$\tilde{\varkappa}^*=\cos\tilde{\alpha}/\cos(2\tilde{\alpha})$,
		\textcolor{black}{$\tilde{\varkappa}^*=\tilde{\alpha}-\pi/2$},
		$\Gamma_2$ is obtained numerically). The dashed cyan curve $\Gamma_3$ and dashed green curve $\Gamma_4$ show the numerically computed stability boundaries for generalized splay states with $r_2=0$ in the delayed system and in the second-order reduced model \eqref{eq:reduced-model-2c}, respectively.} 
	\label{fig:fig1}
	\vspace{-5mm}
\end{figure}

{\it Nonlocal Kuramoto-Battogtokh coupling.} {We next test the reduction on the time-delayed Kuramoto--Battogtokh nonlocal network, a stringent chimera benchmark whose shape is highly sensitive to approximation \cite{kuramoto2002coexistence}.} We consider identical oscillators arranged at equal intervals on a ring segment of length $L$, with positions $x_j=jL/N$. The nonlocal coupling is defined by the exponential kernel \cite{kuramoto2002coexistence,smirnov2017chimera}:
%\begin{equation}
$   G(x)={\kappa \cosh\!\left( \kappa \left|x\right| \right)}\Big/{2 \sinh(\kappa L/2)},$
% \label{7}
%\end{equation}
and the coupling functions 
%\begin{equation}
$F_{jk}\bigl(\phi_{k}(t-\tau)-\phi_{j}(t)\bigr)
= G_{jk}\sin\bigl(\phi_{k}(t-\tau)-\phi_{j}(t)-\alpha\bigr)$ with phase shift $\alpha$ and coupling coefficients $G_{jk}=G(x_k-x_j)$.
Figure~\ref{fig:2-revised}(a--d) compares the dynamics of the delayed Kuramoto–Battogtokh system with its first- and second-order reductions. {For 250 random initial conditions, Fig.~\ref{fig:2-revised}(a-d) shows that the second-order model closely reproduces the delayed system, including the chimera branch, phase profile, synchronized-cluster size, and probabilities of synchrony versus chimera convergence. The first-order reduction gives distorted $r_1$ dynamics and substantially misestimates these probabilities.}
The inset shows that the phase profile of a representative chimera, such as its cluster size and
\begin{figure}[h!]
	\includegraphics[width=0.85\columnwidth]{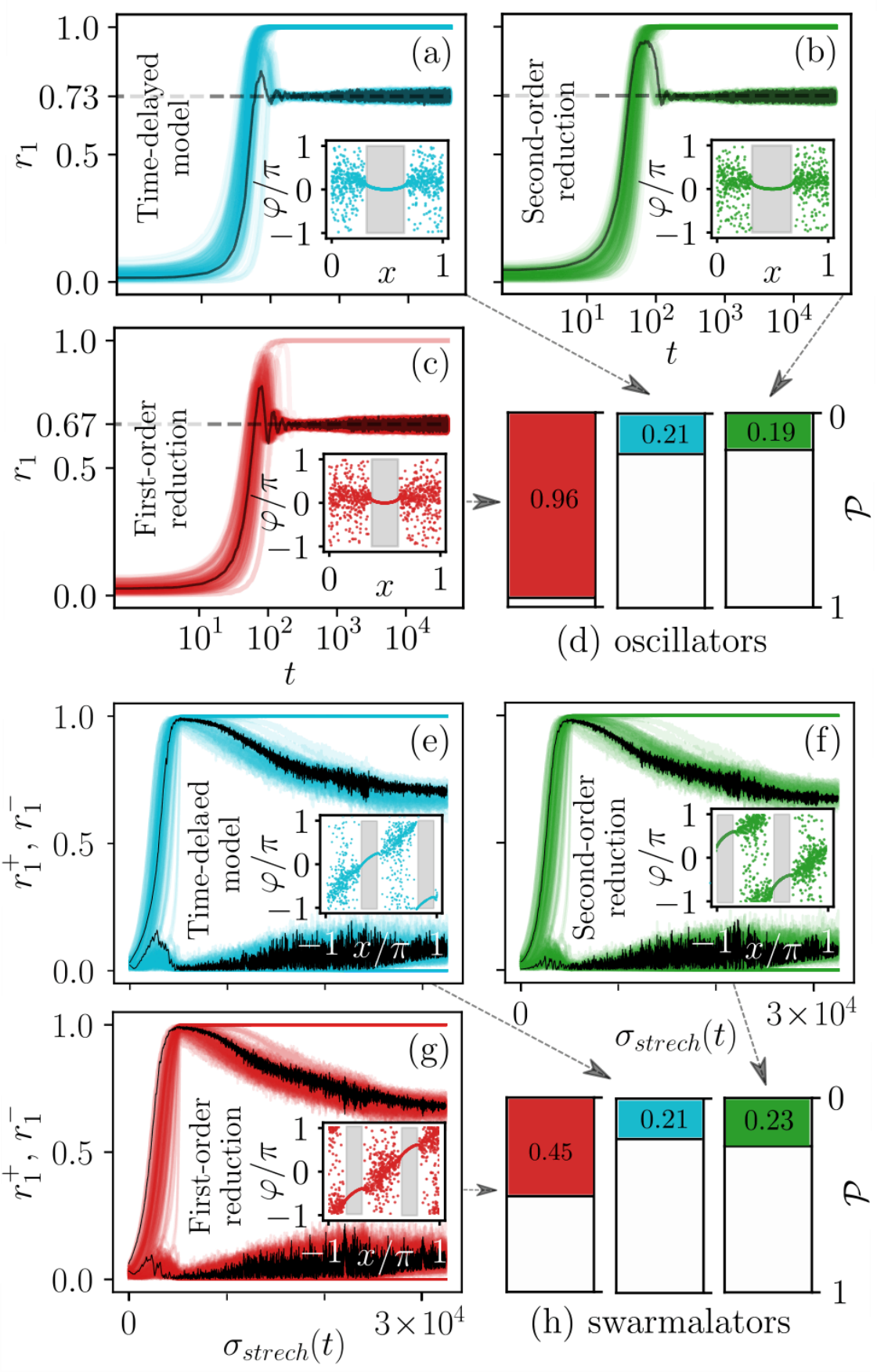}
	\vspace{-3mm}
	\caption{{Comparison of the predictive power of the first- and second-order reductions in two challenging settings: chimera states in the time-delayed Kuramoto--Battogtokh model and hybrid chimera-like states in the time-delayed swarmalator model. Panels (a)--(d) correspond to the Kuramoto--Battogtokh system \eqref{eq:kd-model} with nonlocal coupling: (a) delayed model, (b) second-order reduction \eqref{eq:reduced-model-2c}, and (c) first-order reduction \eqref{eq:reduced-model-1}. Each panel shows results from 250 simulations with random initial phases $\phi_j(0)\in[-\pi,\pi]$ (and zero initial velocities $\dot{\phi}_j(0)=0$ for the second-order model). The black curve highlights a representative chimera trajectory; the inset shows the corresponding phase profile $\varphi_j=\phi_j-\phi_{N/2}$ aligned with the central oscillator, with the gray band marking the synchronized cluster. Panel (d) shows the probabilities $\mathcal{P}$ of convergence to chimera states (colored segment) and to full synchrony (white segment). Parameters: $N=1024$, $L=1.0$, $\kappa=5.2$, $\tau=0.08$, $\varpi=5.125$, $\alpha=1.047$. Panels (e)--(h) show the corresponding comparison for the time-delayed swarmalator model: (e) delayed model, (f) second-order reduction, and (g) corresponding first-order reduction. A time scale $\sigma_{strech}(t)=10t -450\ln\!\left((1+e^{(t-500)/50})/(1+e^{-10})\right)$ is used, which stretches the initial segment by a factor of~10 and smoothly transitions to the normal scale after $t\approx500$. These panels show the dynamics of the space-phase order parameters $r_1^\pm$ (see Supplementary Material for definitions); the black curve again marks a representative hybrid chimera-like trajectory. Panel (h) shows the probabilities $\mathcal{P}$ of convergence to hybrid chimera-like states (colored segment) and to the plane-wave regime (white segment). Parameters: $N=1024$, $L=2\pi$, $\tau=0.05$, $\sigma=0.001$, $\kappa=0.6$, $\beta=1.4135$, $\alpha=0.0625$, $\nu=3.142$, $\omega=0.157$.}}
	\vspace{-12mm}
	\label{fig:2-revised}
\end{figure}
shape of the synchronized domain, matches almost perfectly between the delayed system and the second-order reduction. In contrast, the first-order model produces a visibly different pattern. The probability panels on the right confirm this claim quantitatively: the second-order model captures the coexistence statistics of fully synchronous and chimera states observed in the delayed system, while the first-order reduction both underestimates full synchronization and misidentifies the prevalence of chimera states.

{\it Small-world network.} {We also tested a Watts--Strogatz network \cite{watts1998collective} with 
	$F_{jk}=A_{jk}\sin(\phi_k(t-\tau)-\phi_j(t))$, where $A_{jk}=Na_{jk}/\kappa_j$ and $a_{jk}$ is the randomly rewired adjacency matrix in Fig.~\ref{fig:ReW0eq0}(a). This provides an explicit test on a heterogeneous random network. } This network supports twisted states \cite{acebron,laing2016travelling}, i.e., partially synchronous patterns in which the phases wind nontrivially around the circle and form nonuniform spatial profiles. Figure~\ref{fig:ReW0eq0} shows that the second-order model reproduces the delayed system’s twisted-state statistics, order-parameter dynamics, phase snapshots, and phase-density profiles, whereas the first-order reduction gives distorted statistics and profiles.
\begin{figure}[h!]
	\includegraphics[width=0.9\columnwidth]{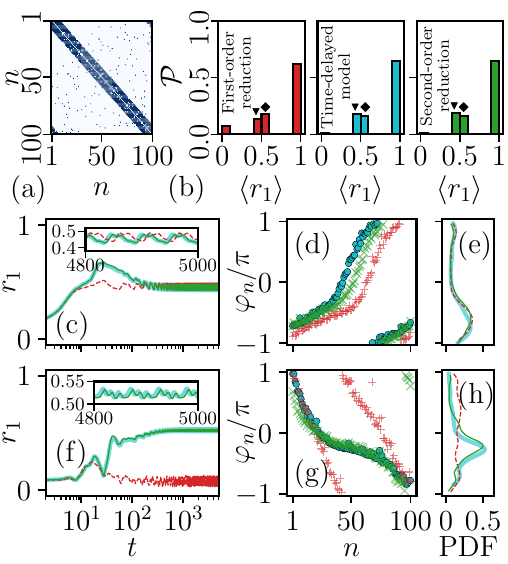}
	\vspace{-4mm}
	\caption{(a) Adjacency matrix of the time-delayed small-world network ($N=100$, mean degree $\langle\kappa\rangle=16$, rewiring probability $p=0.08$). (b) Probabilities $\mathcal{P}$ of realizing different stable regimes for random initial conditions in the delayed network and in the corresponding first- and second-order reduced models (2,000 runs; initial phases drawn as random constants). Triangle and diamond markers indicate bins corresponding to the examples shown in (d) and (g). (c,f) Time series of the order parameter $r_1$ for selected regimes, comparing the delayed system (cyan), the second-order reduction (green), and the first-order reduction (red) under identical initial conditions. (d,g) Instantaneous phase snapshots, and (e,h) phase-density profiles (PDFs), illustrating partially synchronous nonuniform twisted states. Parameters: $\varkappa=0.05$, $\varpi=0.94$, $\tau=1$.
	}
	\vspace{-1mm}
	\label{fig:ReW0eq0}
\end{figure}

{\it Global bi-harmonic coupling.} 
We finally consider global biharmonic coupling,
$
F_{jk}\bigl(\phi_{k}(t-\tau)-\phi_{j}(t)\bigr)
= \sum \limits_{q=1}^{2} K_q \sin\bigl({q\left(\phi_{k}(t-\tau)-\phi_{j}(t)\right)}-\alpha_q\bigr),
$
%\label{eq:fun_GlobCoupl_2H}
%\end{equation}
where $K_q$ and $\alpha_q$ are the strength and phase shift of the $q$-th harmonic. 
The second-order reduction captures cyclops states \cite{munyayev2023cyclops} such as three-cluster generalized splay states with two coherent clusters and one solitary oscillator, and their nonstationary breathing and switching variants \cite{bolotov2024breathing,bolotov2025heterogeneity}. {These states are genuinely two-dimensional features of the inertial reduction and are not captured by the first-order phase-only model.}
Figure~\ref{fig-3} shows that the second-order reduction accurately captures both stationary cyclops states and nonstationary breathing and switching dynamics, tracking the order parameters $r_1,r_2$ and the corresponding phase configurations.

{To probe the validity range of the reduction in this most demanding nonstationary regime, Supplementary Fig.~1 follows breathing cyclops dynamics as $\tau$ increases at fixed $\kappa$, showing that the second-order model retains quantitative accuracy, well beyond the formally asymptotic range, up to $\tau\approx 3.5$ (i.e., $\kappa\tau\approx 0.35$).}
\begin{figure}[h]
	\includegraphics[width=0.9\columnwidth]{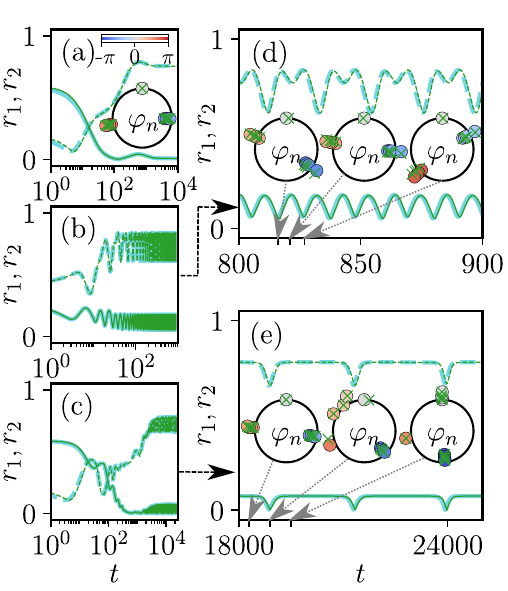}
	\vspace{-1mm}
	\caption{Cyclops, breathing and switching cyclops states in the delayed bi-harmonic KD model (cyan) and its second-order reduction (green). Panels show the time evolution of the first ($r_1$, solid) and second ($r_2$, dashed) Kuramoto order parameters for (a) a stationary cyclops state ($\alpha_1=1.9$, $\alpha_2=-2.4$), (b,d) a breathing cyclops state ($\alpha_1=0.42$, $\alpha_2=-2.3$), and (c,e) a switching cyclops state ($\alpha_1=0.6$, $\alpha_2=-0.92$). Insets in (a,d,e) display instantaneous phase snapshots for the delayed system (colored circles) and the second-order reduction (crosses), illustrating the match of cluster structure and solitary oscillator position. Initial conditions are uniformly random constant phases. Parameters: $N=9$, $\tau=1$, $\varkappa=0.1$, {$K_1=1$, $K_2=0.05$}, $\varpi=1.2$.}\label{fig-3}
	\vspace{-4mm}
\end{figure}
{\paragraph{Beyond phase oscillators: time-delayed swarmalators.}
	The same reduction framework extends beyond phase-only networks to mobile oscillator ensembles with coupled spatial and phase dynamics \cite{Keeffe2017,Yoon2022,Hong2023,Lizarraga2023,Cai2024,Keeffe2026}. As shown in the Supplementary Material, a one-dimensional time-delayed swarmalator model admits an analogous second-order delay-free reduction in which delay again induces effective inertia and triadic couplings, now linking the spatial and phase sectors. To characterize the resulting collective states, we use the space-phase order parameters $r_1^\pm$, which measure coherence in the sum and difference variables of the swarmalator dynamics (see Supplementary Material for definitions and derivation). Figure~\ref{fig:2-revised}(e--h) shows that the reduced model predicts both the emergence and the collective-state statistics of a hybrid chimera-like state with coherent clusters on a plane-wave background remarkably well, with probability $0.21$ versus $0.23$ in the time-delayed system, whereas the first-order reduction substantially overestimates this state, assigning it probability $0.45$. This extension establishes a universal predictive framework for time-delayed oscillator networks, showing that the delay-to-inertia-and-triadic-interactions mechanism applies beyond phase-oscillator systems.}\\
\textit{Conclusions.} We have developed a universal second-order {predictive} reduction for time-delayed KD networks, mapping delayed 1D phase dynamics to a delay-free network of 2D rotators. The reduction converts delay-induced memory into effective inertia and higher-order (triadic) interactions, yielding a compact, parameter-explicit model that {predicts nontrivial attractors and their collective-state statistics, including the emergence probabilities and shapes of splay, chimera, and cyclops regimes}. Across global, nonlocal, small-world, biharmonic, and {swarmalator} examples, it captures complex, often nonstationary collective dynamics beyond the reach of first-order phase reductions, {supporting the universality of the delay-to-inertia-and-triadic-interactions mechanism}. In the weak-coupling regime, the time-delayed 1D Kuramoto model is dynamically equivalent to a delay-free 2D Kuramoto-type model with inertia and, where applicable, higher-order interactions. This mapping provides a new application domain for inertial Kuramoto models, beyond their established roles in adaptive-frequency firefly synchronization \cite{ermentrout1991adaptive}, power-grid oscillator networks \cite{tumash2019stability}, and theta-neuron populations with adaptive synaptic coupling \cite{munyayev2023cyclops,smirnov2024synaptic}. Since many delayed physical and biological systems, including integrate-and-fire neurons \cite{gerstner2002spiking}, Josephson junction arrays \cite{wiesenfeld1996synchronization,wiesenfeld1998frequency,trees2005synchronization}, and laser oscillators with delayed feedback \cite{garbin2015topological,nair2021using}, admit KD-type phase descriptions, the present work provides {a universal predictive framework for analyzing and designing delay-controlled collective behavior in oscillator networks.}\\
{\it Acknowledgments.} We are grateful to G.V. Osipov for valuable discussions. This work was supported by the RSF under Project No. 22-12-00348-P (model formulations and reductions, global coupling models and small-world network), Project No. 24-72-00105 (nonlocal coupling model and swarmalators)
%(to V.O.M., and M.I.B.), the MSHE under Project No. FSWR-2024–0005 (to L.A.S., Supplemental Material),
and the National Science Foundation (USA) under grant DMS-2510860 (to I.B.).

\bibliography{references_from_delay_to_inertia}

\begin{thebibliography}{66}%
\makeatletter
\providecommand \@ifxundefined [1]{%
 \@ifx{#1\undefined}
}%
\providecommand \@ifnum [1]{%
 \ifnum #1\expandafter \@firstoftwo
 \else \expandafter \@secondoftwo
 \fi
}%
\providecommand \@ifx [1]{%
 \ifx #1\expandafter \@firstoftwo
 \else \expandafter \@secondoftwo
 \fi
}%
\providecommand \natexlab [1]{#1}%
\providecommand \enquote  [1]{``#1''}%
\providecommand \bibnamefont  [1]{#1}%
\providecommand \bibfnamefont [1]{#1}%
\providecommand \citenamefont [1]{#1}%
\providecommand \href@noop [0]{\@secondoftwo}%
\providecommand \href [0]{\begingroup \@sanitize@url \@href}%
\providecommand \@href[1]{\@@startlink{#1}\@@href}%
\providecommand \@@href[1]{\endgroup#1\@@endlink}%
\providecommand \@sanitize@url [0]{\catcode `\\12\catcode `\$12\catcode
  `\&12\catcode `\#12\catcode `\^12\catcode `\_12\catcode `\%12\relax}%
\providecommand \@@startlink[1]{}%
\providecommand \@@endlink[0]{}%
\providecommand \url  [0]{\begingroup\@sanitize@url \@url }%
\providecommand \@url [1]{\endgroup\@href {#1}{\urlprefix }}%
\providecommand \urlprefix  [0]{URL }%
\providecommand \Eprint [0]{\href }%
\providecommand \doibase [0]{https://doi.org/}%
\providecommand \selectlanguage [0]{\@gobble}%
\providecommand \bibinfo  [0]{\@secondoftwo}%
\providecommand \bibfield  [0]{\@secondoftwo}%
\providecommand \translation [1]{[#1]}%
\providecommand \BibitemOpen [0]{}%
\providecommand \bibitemStop [0]{}%
\providecommand \bibitemNoStop [0]{.\EOS\space}%
\providecommand \EOS [0]{\spacefactor3000\relax}%
\providecommand \BibitemShut  [1]{\csname bibitem#1\endcsname}%
\let\auto@bib@innerbib\@empty
%</preamble>
\bibitem [{\citenamefont {Izhikevich}(1998)}]{izhikevich1998phase}%
  \BibitemOpen
  \bibfield  {author} {\bibinfo {author} {\bibfnamefont {E.~M.}\ \bibnamefont
  {Izhikevich}},\ }\bibfield  {title} {\bibinfo {title} {Phase models with
  explicit time delays},\ }\href@noop {} {\bibfield  {journal} {\bibinfo
  {journal} {Physical Review E}\ }\textbf {\bibinfo {volume} {58}},\ \bibinfo
  {pages} {905} (\bibinfo {year} {1998})}\BibitemShut {NoStop}%
\bibitem [{\citenamefont {Yeung}\ and\ \citenamefont
  {Strogatz}(1999)}]{yeung1999time}%
  \BibitemOpen
  \bibfield  {author} {\bibinfo {author} {\bibfnamefont {M.~S.}\ \bibnamefont
  {Yeung}}\ and\ \bibinfo {author} {\bibfnamefont {S.~H.}\ \bibnamefont
  {Strogatz}},\ }\bibfield  {title} {\bibinfo {title} {Time delay in the
  kuramoto model of coupled oscillators},\ }\href@noop {} {\bibfield  {journal}
  {\bibinfo  {journal} {Physical Review Letters}\ }\textbf {\bibinfo {volume}
  {82}},\ \bibinfo {pages} {648} (\bibinfo {year} {1999})}\BibitemShut
  {NoStop}%
\bibitem [{\citenamefont {Earl}\ and\ \citenamefont
  {Strogatz}(2003{\natexlab{a}})}]{earl2003synchronization}%
  \BibitemOpen
  \bibfield  {author} {\bibinfo {author} {\bibfnamefont {M.~G.}\ \bibnamefont
  {Earl}}\ and\ \bibinfo {author} {\bibfnamefont {S.~H.}\ \bibnamefont
  {Strogatz}},\ }\bibfield  {title} {\bibinfo {title} {Synchronization in
  oscillator networks with delayed coupling: A stability criterion},\
  }\href@noop {} {\bibfield  {journal} {\bibinfo  {journal} {Physical Review
  E}\ }\textbf {\bibinfo {volume} {67}},\ \bibinfo {pages} {036204} (\bibinfo
  {year} {2003}{\natexlab{a}})}\BibitemShut {NoStop}%
\bibitem [{\citenamefont {J{\"o}rg}\ \emph {et~al.}(2014)\citenamefont
  {J{\"o}rg}, \citenamefont {Morelli}, \citenamefont {Ares},\ and\
  \citenamefont {J{\"u}licher}}]{jorg2014synchronization}%
  \BibitemOpen
  \bibfield  {author} {\bibinfo {author} {\bibfnamefont {D.~J.}\ \bibnamefont
  {J{\"o}rg}}, \bibinfo {author} {\bibfnamefont {L.~G.}\ \bibnamefont
  {Morelli}}, \bibinfo {author} {\bibfnamefont {S.}~\bibnamefont {Ares}},\ and\
  \bibinfo {author} {\bibfnamefont {F.}~\bibnamefont {J{\"u}licher}},\
  }\bibfield  {title} {\bibinfo {title} {Synchronization dynamics in the
  presence of coupling delays and phase shifts},\ }\href@noop {} {\bibfield
  {journal} {\bibinfo  {journal} {Physical Review Letters}\ }\textbf {\bibinfo
  {volume} {112}},\ \bibinfo {pages} {174101} (\bibinfo {year}
  {2014})}\BibitemShut {NoStop}%
\bibitem [{\citenamefont {Restrepo}\ and\ \citenamefont
  {Skardal}(2019)}]{restrepo2019competitive}%
  \BibitemOpen
  \bibfield  {author} {\bibinfo {author} {\bibfnamefont {J.~G.}\ \bibnamefont
  {Restrepo}}\ and\ \bibinfo {author} {\bibfnamefont {P.~S.}\ \bibnamefont
  {Skardal}},\ }\bibfield  {title} {\bibinfo {title} {Competitive suppression
  of synchronization and nonmonotonic transitions in oscillator communities
  with distributed time delay},\ }\href@noop {} {\bibfield  {journal} {\bibinfo
   {journal} {Physical Review Research}\ }\textbf {\bibinfo {volume} {1}},\
  \bibinfo {pages} {033042} (\bibinfo {year} {2019})}\BibitemShut {NoStop}%
\bibitem [{\citenamefont {Ocampo-Espindola}\ \emph {et~al.}(2024)\citenamefont
  {Ocampo-Espindola}, \citenamefont {Kiss}, \citenamefont {Bick},\ and\
  \citenamefont {Wedgwood}}]{ocampo2024strong}%
  \BibitemOpen
  \bibfield  {author} {\bibinfo {author} {\bibfnamefont {J.~L.}\ \bibnamefont
  {Ocampo-Espindola}}, \bibinfo {author} {\bibfnamefont {I.~Z.}\ \bibnamefont
  {Kiss}}, \bibinfo {author} {\bibfnamefont {C.}~\bibnamefont {Bick}},\ and\
  \bibinfo {author} {\bibfnamefont {K.~C.}\ \bibnamefont {Wedgwood}},\
  }\bibfield  {title} {\bibinfo {title} {Strong coupling yields abrupt
  synchronization transitions in coupled oscillators},\ }\href@noop {}
  {\bibfield  {journal} {\bibinfo  {journal} {Physical Review Research}\
  }\textbf {\bibinfo {volume} {6}},\ \bibinfo {pages} {033328} (\bibinfo {year}
  {2024})}\BibitemShut {NoStop}%
\bibitem [{\citenamefont {Bick}\ \emph {et~al.}(2024)\citenamefont {Bick},
  \citenamefont {Rink},\ and\ \citenamefont {de~Wolff}}]{bick2024time}%
  \BibitemOpen
  \bibfield  {author} {\bibinfo {author} {\bibfnamefont {C.}~\bibnamefont
  {Bick}}, \bibinfo {author} {\bibfnamefont {B.}~\bibnamefont {Rink}},\ and\
  \bibinfo {author} {\bibfnamefont {B.~A.}\ \bibnamefont {de~Wolff}},\
  }\bibfield  {title} {\bibinfo {title} {When time delays and phase lags are
  not the same: higher-order phase reduction unravels delay-induced
  synchronization in oscillator networks},\ }\href@noop {} {\bibfield
  {journal} {\bibinfo  {journal} {arXiv preprint arXiv:2404.11340}\ } (\bibinfo
  {year} {2024})}\BibitemShut {NoStop}%
\bibitem [{\citenamefont {Lee}\ \emph {et~al.}(2009)\citenamefont {Lee},
  \citenamefont {Ott},\ and\ \citenamefont {Antonsen}}]{lee2009large}%
  \BibitemOpen
  \bibfield  {author} {\bibinfo {author} {\bibfnamefont {W.~S.}\ \bibnamefont
  {Lee}}, \bibinfo {author} {\bibfnamefont {E.}~\bibnamefont {Ott}},\ and\
  \bibinfo {author} {\bibfnamefont {T.~M.}\ \bibnamefont {Antonsen}},\
  }\bibfield  {title} {\bibinfo {title} {Large coupled oscillator systems with
  heterogeneous interaction delays},\ }\href@noop {} {\bibfield  {journal}
  {\bibinfo  {journal} {Physical Review Letters}\ }\textbf {\bibinfo {volume}
  {103}},\ \bibinfo {pages} {044101} (\bibinfo {year} {2009})}\BibitemShut
  {NoStop}%
\bibitem [{\citenamefont {Bick}\ \emph {et~al.}(2017)\citenamefont {Bick},
  \citenamefont {Sebek},\ and\ \citenamefont {Kiss}}]{bick2017robust}%
  \BibitemOpen
  \bibfield  {author} {\bibinfo {author} {\bibfnamefont {C.}~\bibnamefont
  {Bick}}, \bibinfo {author} {\bibfnamefont {M.}~\bibnamefont {Sebek}},\ and\
  \bibinfo {author} {\bibfnamefont {I.~Z.}\ \bibnamefont {Kiss}},\ }\bibfield
  {title} {\bibinfo {title} {Robust weak chimeras in oscillator networks with
  delayed linear and quadratic interactions},\ }\href@noop {} {\bibfield
  {journal} {\bibinfo  {journal} {Physical Review Letters}\ }\textbf {\bibinfo
  {volume} {119}},\ \bibinfo {pages} {168301} (\bibinfo {year}
  {2017})}\BibitemShut {NoStop}%
\bibitem [{\citenamefont {Ameli}\ \emph {et~al.}(2021)\citenamefont {Ameli},
  \citenamefont {Karimian},\ and\ \citenamefont {Shahbazi}}]{ameli2021time}%
  \BibitemOpen
  \bibfield  {author} {\bibinfo {author} {\bibfnamefont {S.}~\bibnamefont
  {Ameli}}, \bibinfo {author} {\bibfnamefont {M.}~\bibnamefont {Karimian}},\
  and\ \bibinfo {author} {\bibfnamefont {F.}~\bibnamefont {Shahbazi}},\
  }\bibfield  {title} {\bibinfo {title} {Time-delayed kuramoto model in the
  watts--strogatz small-world networks},\ }\href@noop {} {\bibfield  {journal}
  {\bibinfo  {journal} {Chaos: An Interdisciplinary Journal of Nonlinear
  Science}\ }\textbf {\bibinfo {volume} {31}} (\bibinfo {year}
  {2021})}\BibitemShut {NoStop}%
\bibitem [{\citenamefont {Gjurchinovski}\ \emph {et~al.}(2017)\citenamefont
  {Gjurchinovski}, \citenamefont {Sch{\"o}ll},\ and\ \citenamefont
  {Zakharova}}]{gjurchinovski2017control}%
  \BibitemOpen
  \bibfield  {author} {\bibinfo {author} {\bibfnamefont {A.}~\bibnamefont
  {Gjurchinovski}}, \bibinfo {author} {\bibfnamefont {E.}~\bibnamefont
  {Sch{\"o}ll}},\ and\ \bibinfo {author} {\bibfnamefont {A.}~\bibnamefont
  {Zakharova}},\ }\bibfield  {title} {\bibinfo {title} {Control of amplitude
  chimeras by time delay in oscillator networks},\ }\href@noop {} {\bibfield
  {journal} {\bibinfo  {journal} {Physical Review E}\ }\textbf {\bibinfo
  {volume} {95}},\ \bibinfo {pages} {042218} (\bibinfo {year}
  {2017})}\BibitemShut {NoStop}%
\bibitem [{\citenamefont {Sheeba}\ \emph {et~al.}(2009)\citenamefont {Sheeba},
  \citenamefont {Chandrasekar},\ and\ \citenamefont
  {Lakshmanan}}]{sheeba2009globally}%
  \BibitemOpen
  \bibfield  {author} {\bibinfo {author} {\bibfnamefont {J.~H.}\ \bibnamefont
  {Sheeba}}, \bibinfo {author} {\bibfnamefont {V.}~\bibnamefont
  {Chandrasekar}},\ and\ \bibinfo {author} {\bibfnamefont {M.}~\bibnamefont
  {Lakshmanan}},\ }\bibfield  {title} {\bibinfo {title} {Globally clustered
  chimera states in delay-coupled populations},\ }\href@noop {} {\bibfield
  {journal} {\bibinfo  {journal} {Physical Review E—Statistical, Nonlinear,
  and Soft Matter Physics}\ }\textbf {\bibinfo {volume} {79}},\ \bibinfo
  {pages} {055203} (\bibinfo {year} {2009})}\BibitemShut {NoStop}%
\bibitem [{\citenamefont {Lee}\ \emph {et~al.}(2011)\citenamefont {Lee},
  \citenamefont {Restrepo}, \citenamefont {Ott},\ and\ \citenamefont
  {Antonsen}}]{lee2011dynamics}%
  \BibitemOpen
  \bibfield  {author} {\bibinfo {author} {\bibfnamefont {W.~S.}\ \bibnamefont
  {Lee}}, \bibinfo {author} {\bibfnamefont {J.~G.}\ \bibnamefont {Restrepo}},
  \bibinfo {author} {\bibfnamefont {E.}~\bibnamefont {Ott}},\ and\ \bibinfo
  {author} {\bibfnamefont {T.~M.}\ \bibnamefont {Antonsen}},\ }\bibfield
  {title} {\bibinfo {title} {Dynamics and pattern formation in large systems of
  spatially-coupled oscillators with finite response times},\ }\href@noop {}
  {\bibfield  {journal} {\bibinfo  {journal} {Chaos: An Interdisciplinary
  Journal of Nonlinear Science}\ }\textbf {\bibinfo {volume} {21}},\ \bibinfo
  {pages} {023122} (\bibinfo {year} {2011})}\BibitemShut {NoStop}%
\bibitem [{\citenamefont {Laing}(2016)}]{laing2016travelling}%
  \BibitemOpen
  \bibfield  {author} {\bibinfo {author} {\bibfnamefont {C.~R.}\ \bibnamefont
  {Laing}},\ }\bibfield  {title} {\bibinfo {title} {Travelling waves in arrays
  of delay-coupled phase oscillators},\ }\href@noop {} {\bibfield  {journal}
  {\bibinfo  {journal} {Chaos: An Interdisciplinary Journal of Nonlinear
  Science}\ }\textbf {\bibinfo {volume} {26}},\ \bibinfo {pages} {094802}
  (\bibinfo {year} {2016})}\BibitemShut {NoStop}%
\bibitem [{\citenamefont {An}\ \emph {et~al.}(2024)\citenamefont {An},
  \citenamefont {Ho}, \citenamefont {Kim},\ and\ \citenamefont
  {Choe}}]{an2024stability}%
  \BibitemOpen
  \bibfield  {author} {\bibinfo {author} {\bibfnamefont {Y.-H.}\ \bibnamefont
  {An}}, \bibinfo {author} {\bibfnamefont {M.-S.}\ \bibnamefont {Ho}}, \bibinfo
  {author} {\bibfnamefont {R.-S.}\ \bibnamefont {Kim}},\ and\ \bibinfo {author}
  {\bibfnamefont {C.-U.}\ \bibnamefont {Choe}},\ }\bibfield  {title} {\bibinfo
  {title} {Stability of the twisted states in a ring of oscillators interacting
  with distance-dependent delays},\ }\href@noop {} {\bibfield  {journal}
  {\bibinfo  {journal} {Physica D: Nonlinear Phenomena}\ }\textbf {\bibinfo
  {volume} {464}},\ \bibinfo {pages} {134204} (\bibinfo {year}
  {2024})}\BibitemShut {NoStop}%
\bibitem [{\citenamefont {Sawicki}\ \emph {et~al.}(2017)\citenamefont
  {Sawicki}, \citenamefont {Omelchenko}, \citenamefont {Zakharova},\ and\
  \citenamefont {Sch{\"o}ll}}]{sawicki2017chimera}%
  \BibitemOpen
  \bibfield  {author} {\bibinfo {author} {\bibfnamefont {J.}~\bibnamefont
  {Sawicki}}, \bibinfo {author} {\bibfnamefont {I.}~\bibnamefont {Omelchenko}},
  \bibinfo {author} {\bibfnamefont {A.}~\bibnamefont {Zakharova}},\ and\
  \bibinfo {author} {\bibfnamefont {E.}~\bibnamefont {Sch{\"o}ll}},\ }\bibfield
   {title} {\bibinfo {title} {Chimera states in complex networks: interplay of
  fractal topology and delay},\ }\href@noop {} {\bibfield  {journal} {\bibinfo
  {journal} {The European Physical Journal Special Topics}\ }\textbf {\bibinfo
  {volume} {226}},\ \bibinfo {pages} {1883} (\bibinfo {year}
  {2017})}\BibitemShut {NoStop}%
\bibitem [{\citenamefont {Wu}\ and\ \citenamefont
  {Dhamala}(2018)}]{wu2018dynamics}%
  \BibitemOpen
  \bibfield  {author} {\bibinfo {author} {\bibfnamefont {H.}~\bibnamefont
  {Wu}}\ and\ \bibinfo {author} {\bibfnamefont {M.}~\bibnamefont {Dhamala}},\
  }\bibfield  {title} {\bibinfo {title} {Dynamics of kuramoto oscillators with
  time-delayed positive and negative couplings},\ }\href@noop {} {\bibfield
  {journal} {\bibinfo  {journal} {Physical Review E}\ }\textbf {\bibinfo
  {volume} {98}},\ \bibinfo {pages} {032221} (\bibinfo {year}
  {2018})}\BibitemShut {NoStop}%
\bibitem [{\citenamefont {Wolfrum}\ and\ \citenamefont
  {Yanchuk}(2006)}]{wolfrum2006eckhaus}%
  \BibitemOpen
  \bibfield  {author} {\bibinfo {author} {\bibfnamefont {M.}~\bibnamefont
  {Wolfrum}}\ and\ \bibinfo {author} {\bibfnamefont {S.}~\bibnamefont
  {Yanchuk}},\ }\bibfield  {title} {\bibinfo {title} {Eckhaus instability in
  systems with large delay},\ }\href@noop {} {\bibfield  {journal} {\bibinfo
  {journal} {Physical Review Letters}\ }\textbf {\bibinfo {volume} {96}},\
  \bibinfo {pages} {220201} (\bibinfo {year} {2006})}\BibitemShut {NoStop}%
\bibitem [{\citenamefont {Sethia}\ \emph {et~al.}(2008)\citenamefont {Sethia},
  \citenamefont {Sen},\ and\ \citenamefont {Atay}}]{sethia2008clustered}%
  \BibitemOpen
  \bibfield  {author} {\bibinfo {author} {\bibfnamefont {G.~C.}\ \bibnamefont
  {Sethia}}, \bibinfo {author} {\bibfnamefont {A.}~\bibnamefont {Sen}},\ and\
  \bibinfo {author} {\bibfnamefont {F.~M.}\ \bibnamefont {Atay}},\ }\bibfield
  {title} {\bibinfo {title} {Clustered chimera states in delay-coupled
  oscillator systems},\ }\href@noop {} {\bibfield  {journal} {\bibinfo
  {journal} {Physical Review Letters}\ }\textbf {\bibinfo {volume} {100}},\
  \bibinfo {pages} {144102} (\bibinfo {year} {2008})}\BibitemShut {NoStop}%
\bibitem [{\citenamefont {Ares}\ \emph {et~al.}(2012)\citenamefont {Ares},
  \citenamefont {Morelli}, \citenamefont {J{\"o}rg}, \citenamefont {Oates},\
  and\ \citenamefont {J{\"u}licher}}]{ares2012collective}%
  \BibitemOpen
  \bibfield  {author} {\bibinfo {author} {\bibfnamefont {S.}~\bibnamefont
  {Ares}}, \bibinfo {author} {\bibfnamefont {L.~G.}\ \bibnamefont {Morelli}},
  \bibinfo {author} {\bibfnamefont {D.~J.}\ \bibnamefont {J{\"o}rg}}, \bibinfo
  {author} {\bibfnamefont {A.~C.}\ \bibnamefont {Oates}},\ and\ \bibinfo
  {author} {\bibfnamefont {F.}~\bibnamefont {J{\"u}licher}},\ }\bibfield
  {title} {\bibinfo {title} {Collective modes of coupled phase oscillators with
  delayed coupling},\ }\href@noop {} {\bibfield  {journal} {\bibinfo  {journal}
  {Physical Review Letters}\ }\textbf {\bibinfo {volume} {108}},\ \bibinfo
  {pages} {204101} (\bibinfo {year} {2012})}\BibitemShut {NoStop}%
\bibitem [{\citenamefont {Yanchuk}\ and\ \citenamefont
  {Giacomelli}(2014)}]{yanchuk2014pattern}%
  \BibitemOpen
  \bibfield  {author} {\bibinfo {author} {\bibfnamefont {S.}~\bibnamefont
  {Yanchuk}}\ and\ \bibinfo {author} {\bibfnamefont {G.}~\bibnamefont
  {Giacomelli}},\ }\bibfield  {title} {\bibinfo {title} {Pattern formation in
  systems with multiple delayed feedbacks},\ }\href@noop {} {\bibfield
  {journal} {\bibinfo  {journal} {Physical Review Letters}\ }\textbf {\bibinfo
  {volume} {112}},\ \bibinfo {pages} {174103} (\bibinfo {year}
  {2014})}\BibitemShut {NoStop}%
\bibitem [{\citenamefont {Dhamala}\ \emph {et~al.}(2004)\citenamefont
  {Dhamala}, \citenamefont {Jirsa},\ and\ \citenamefont
  {Ding}}]{dhamala2004enhancement}%
  \BibitemOpen
  \bibfield  {author} {\bibinfo {author} {\bibfnamefont {M.}~\bibnamefont
  {Dhamala}}, \bibinfo {author} {\bibfnamefont {V.~K.}\ \bibnamefont {Jirsa}},\
  and\ \bibinfo {author} {\bibfnamefont {M.}~\bibnamefont {Ding}},\ }\bibfield
  {title} {\bibinfo {title} {Enhancement of neural synchrony by time delay},\
  }\href@noop {} {\bibfield  {journal} {\bibinfo  {journal} {Physical Review
  Letters}\ }\textbf {\bibinfo {volume} {92}},\ \bibinfo {pages} {074104}
  (\bibinfo {year} {2004})}\BibitemShut {NoStop}%
\bibitem [{\citenamefont {Sun}\ \emph {et~al.}(2017)\citenamefont {Sun},
  \citenamefont {Perc},\ and\ \citenamefont {Kurths}}]{sun2017effects}%
  \BibitemOpen
  \bibfield  {author} {\bibinfo {author} {\bibfnamefont {X.}~\bibnamefont
  {Sun}}, \bibinfo {author} {\bibfnamefont {M.}~\bibnamefont {Perc}},\ and\
  \bibinfo {author} {\bibfnamefont {J.}~\bibnamefont {Kurths}},\ }\bibfield
  {title} {\bibinfo {title} {Effects of partial time delays on phase
  synchronization in watts-strogatz small-world neuronal networks},\
  }\href@noop {} {\bibfield  {journal} {\bibinfo  {journal} {Chaos: An
  Interdisciplinary Journal of Nonlinear Science}\ }\textbf {\bibinfo {volume}
  {27}},\ \bibinfo {pages} {053113} (\bibinfo {year} {2017})}\BibitemShut
  {NoStop}%
\bibitem [{\citenamefont {Lehnert}\ \emph {et~al.}(2011)\citenamefont
  {Lehnert}, \citenamefont {Dahms}, \citenamefont {H{\"o}vel},\ and\
  \citenamefont {Sch{\"o}ll}}]{lehnert2011loss}%
  \BibitemOpen
  \bibfield  {author} {\bibinfo {author} {\bibfnamefont {J.}~\bibnamefont
  {Lehnert}}, \bibinfo {author} {\bibfnamefont {T.}~\bibnamefont {Dahms}},
  \bibinfo {author} {\bibfnamefont {P.}~\bibnamefont {H{\"o}vel}},\ and\
  \bibinfo {author} {\bibfnamefont {E.}~\bibnamefont {Sch{\"o}ll}},\ }\bibfield
   {title} {\bibinfo {title} {Loss of synchronization in complex neuronal
  networks with delay},\ }\href@noop {} {\bibfield  {journal} {\bibinfo
  {journal} {Europhysics Letters}\ }\textbf {\bibinfo {volume} {96}},\ \bibinfo
  {pages} {60013} (\bibinfo {year} {2011})}\BibitemShut {NoStop}%
\bibitem [{\citenamefont {Popovych}\ \emph {et~al.}(2011)\citenamefont
  {Popovych}, \citenamefont {Yanchuk},\ and\ \citenamefont
  {Tass}}]{popovych2011delay}%
  \BibitemOpen
  \bibfield  {author} {\bibinfo {author} {\bibfnamefont {O.~V.}\ \bibnamefont
  {Popovych}}, \bibinfo {author} {\bibfnamefont {S.}~\bibnamefont {Yanchuk}},\
  and\ \bibinfo {author} {\bibfnamefont {P.~A.}\ \bibnamefont {Tass}},\
  }\bibfield  {title} {\bibinfo {title} {Delay-and coupling-induced firing
  patterns in oscillatory neural loops},\ }\href@noop {} {\bibfield  {journal}
  {\bibinfo  {journal} {Physical Review Letters}\ }\textbf {\bibinfo {volume}
  {107}},\ \bibinfo {pages} {228102} (\bibinfo {year} {2011})}\BibitemShut
  {NoStop}%
\bibitem [{\citenamefont {Keane}\ \emph {et~al.}(2012)\citenamefont {Keane},
  \citenamefont {Dahms}, \citenamefont {Lehnert}, \citenamefont
  {Suryanarayana}, \citenamefont {H{\"o}vel},\ and\ \citenamefont
  {Sch{\"o}ll}}]{keane2012synchronisation}%
  \BibitemOpen
  \bibfield  {author} {\bibinfo {author} {\bibfnamefont {A.}~\bibnamefont
  {Keane}}, \bibinfo {author} {\bibfnamefont {T.}~\bibnamefont {Dahms}},
  \bibinfo {author} {\bibfnamefont {J.}~\bibnamefont {Lehnert}}, \bibinfo
  {author} {\bibfnamefont {S.~A.}\ \bibnamefont {Suryanarayana}}, \bibinfo
  {author} {\bibfnamefont {P.}~\bibnamefont {H{\"o}vel}},\ and\ \bibinfo
  {author} {\bibfnamefont {E.}~\bibnamefont {Sch{\"o}ll}},\ }\bibfield  {title}
  {\bibinfo {title} {Synchronisation in networks of delay-coupled type-i
  excitable systems},\ }\href@noop {} {\bibfield  {journal} {\bibinfo
  {journal} {The European Physical Journal B}\ }\textbf {\bibinfo {volume}
  {85}},\ \bibinfo {pages} {407} (\bibinfo {year} {2012})}\BibitemShut
  {NoStop}%
\bibitem [{\citenamefont {Sawicki}\ \emph
  {et~al.}(2019{\natexlab{a}})\citenamefont {Sawicki}, \citenamefont
  {Omelchenko}, \citenamefont {Zakharova},\ and\ \citenamefont
  {Sch{\"o}ll}}]{sawicki2019delay}%
  \BibitemOpen
  \bibfield  {author} {\bibinfo {author} {\bibfnamefont {J.}~\bibnamefont
  {Sawicki}}, \bibinfo {author} {\bibfnamefont {I.}~\bibnamefont {Omelchenko}},
  \bibinfo {author} {\bibfnamefont {A.}~\bibnamefont {Zakharova}},\ and\
  \bibinfo {author} {\bibfnamefont {E.}~\bibnamefont {Sch{\"o}ll}},\ }\bibfield
   {title} {\bibinfo {title} {Delay-induced chimeras in neural networks with
  fractal topology},\ }\href@noop {} {\bibfield  {journal} {\bibinfo  {journal}
  {The European Physical Journal B}\ }\textbf {\bibinfo {volume} {92}},\
  \bibinfo {pages} {54} (\bibinfo {year} {2019}{\natexlab{a}})}\BibitemShut
  {NoStop}%
\bibitem [{\citenamefont {Sawicki}\ \emph
  {et~al.}(2019{\natexlab{b}})\citenamefont {Sawicki}, \citenamefont {Ghosh},
  \citenamefont {Jalan},\ and\ \citenamefont
  {Zakharova}}]{sawicki2019chimeras}%
  \BibitemOpen
  \bibfield  {author} {\bibinfo {author} {\bibfnamefont {J.}~\bibnamefont
  {Sawicki}}, \bibinfo {author} {\bibfnamefont {S.}~\bibnamefont {Ghosh}},
  \bibinfo {author} {\bibfnamefont {S.}~\bibnamefont {Jalan}},\ and\ \bibinfo
  {author} {\bibfnamefont {A.}~\bibnamefont {Zakharova}},\ }\bibfield  {title}
  {\bibinfo {title} {Chimeras in multiplex networks: interplay of inter-and
  intra-layer delays},\ }\href@noop {} {\bibfield  {journal} {\bibinfo
  {journal} {Frontiers in Applied Mathematics and Statistics}\ }\textbf
  {\bibinfo {volume} {5}},\ \bibinfo {pages} {19} (\bibinfo {year}
  {2019}{\natexlab{b}})}\BibitemShut {NoStop}%
\bibitem [{\citenamefont {Lucchetti}\ \emph {et~al.}(2021)\citenamefont
  {Lucchetti}, \citenamefont {Jensen},\ and\ \citenamefont
  {Heltberg}}]{lucchetti2021emergence}%
  \BibitemOpen
  \bibfield  {author} {\bibinfo {author} {\bibfnamefont {A.}~\bibnamefont
  {Lucchetti}}, \bibinfo {author} {\bibfnamefont {M.~H.}\ \bibnamefont
  {Jensen}},\ and\ \bibinfo {author} {\bibfnamefont {M.~L.}\ \bibnamefont
  {Heltberg}},\ }\bibfield  {title} {\bibinfo {title} {Emergence of chimera
  states in a neuronal model of delayed oscillators},\ }\href@noop {}
  {\bibfield  {journal} {\bibinfo  {journal} {Physical Review Research}\
  }\textbf {\bibinfo {volume} {3}},\ \bibinfo {pages} {033041} (\bibinfo {year}
  {2021})}\BibitemShut {NoStop}%
\bibitem [{\citenamefont {Smirnov}\ \emph {et~al.}(2024)\citenamefont
  {Smirnov}, \citenamefont {Munyayev}, \citenamefont {Bolotov}, \citenamefont
  {Osipov},\ and\ \citenamefont {Belykh}}]{smirnov2024synaptic}%
  \BibitemOpen
  \bibfield  {author} {\bibinfo {author} {\bibfnamefont {L.~A.}\ \bibnamefont
  {Smirnov}}, \bibinfo {author} {\bibfnamefont {V.~O.}\ \bibnamefont
  {Munyayev}}, \bibinfo {author} {\bibfnamefont {M.~I.}\ \bibnamefont
  {Bolotov}}, \bibinfo {author} {\bibfnamefont {G.~V.}\ \bibnamefont
  {Osipov}},\ and\ \bibinfo {author} {\bibfnamefont {I.}~\bibnamefont
  {Belykh}},\ }\bibfield  {title} {\bibinfo {title} {How synaptic function
  controls critical transitions in spiking neuron networks: insight from a
  kuramoto model reduction},\ }\href@noop {} {\bibfield  {journal} {\bibinfo
  {journal} {Frontiers in Network Physiology}\ }\textbf {\bibinfo {volume}
  {4}},\ \bibinfo {pages} {1423023} (\bibinfo {year} {2024})}\BibitemShut
  {NoStop}%
\bibitem [{\citenamefont {Kozyreff}\ \emph {et~al.}(2000)\citenamefont
  {Kozyreff}, \citenamefont {Vladimirov},\ and\ \citenamefont
  {Mandel}}]{kozyreff2000global}%
  \BibitemOpen
  \bibfield  {author} {\bibinfo {author} {\bibfnamefont {G.}~\bibnamefont
  {Kozyreff}}, \bibinfo {author} {\bibfnamefont {A.}~\bibnamefont
  {Vladimirov}},\ and\ \bibinfo {author} {\bibfnamefont {P.}~\bibnamefont
  {Mandel}},\ }\bibfield  {title} {\bibinfo {title} {Global coupling with time
  delay in an array of semiconductor lasers},\ }\href@noop {} {\bibfield
  {journal} {\bibinfo  {journal} {Physical Review Letters}\ }\textbf {\bibinfo
  {volume} {85}},\ \bibinfo {pages} {3809} (\bibinfo {year}
  {2000})}\BibitemShut {NoStop}%
\bibitem [{\citenamefont {T{\"o}pfer}\ \emph {et~al.}(2020)\citenamefont
  {T{\"o}pfer}, \citenamefont {Sigurdsson}, \citenamefont {Pickup},\ and\
  \citenamefont {Lagoudakis}}]{topfer2020time}%
  \BibitemOpen
  \bibfield  {author} {\bibinfo {author} {\bibfnamefont {J.~D.}\ \bibnamefont
  {T{\"o}pfer}}, \bibinfo {author} {\bibfnamefont {H.}~\bibnamefont
  {Sigurdsson}}, \bibinfo {author} {\bibfnamefont {L.}~\bibnamefont {Pickup}},\
  and\ \bibinfo {author} {\bibfnamefont {P.~G.}\ \bibnamefont {Lagoudakis}},\
  }\bibfield  {title} {\bibinfo {title} {Time-delay polaritonics},\ }\href@noop
  {} {\bibfield  {journal} {\bibinfo  {journal} {Communications Physics}\
  }\textbf {\bibinfo {volume} {3}},\ \bibinfo {pages} {2} (\bibinfo {year}
  {2020})}\BibitemShut {NoStop}%
\bibitem [{\citenamefont {Heiligenthal}\ \emph {et~al.}(2011)\citenamefont
  {Heiligenthal}, \citenamefont {Dahms}, \citenamefont {Yanchuk}, \citenamefont
  {J{\"u}ngling}, \citenamefont {Flunkert}, \citenamefont {Kanter},
  \citenamefont {Sch{\"o}ll},\ and\ \citenamefont
  {Kinzel}}]{heiligenthal2011strong}%
  \BibitemOpen
  \bibfield  {author} {\bibinfo {author} {\bibfnamefont {S.}~\bibnamefont
  {Heiligenthal}}, \bibinfo {author} {\bibfnamefont {T.}~\bibnamefont {Dahms}},
  \bibinfo {author} {\bibfnamefont {S.}~\bibnamefont {Yanchuk}}, \bibinfo
  {author} {\bibfnamefont {T.}~\bibnamefont {J{\"u}ngling}}, \bibinfo {author}
  {\bibfnamefont {V.}~\bibnamefont {Flunkert}}, \bibinfo {author}
  {\bibfnamefont {I.}~\bibnamefont {Kanter}}, \bibinfo {author} {\bibfnamefont
  {E.}~\bibnamefont {Sch{\"o}ll}},\ and\ \bibinfo {author} {\bibfnamefont
  {W.}~\bibnamefont {Kinzel}},\ }\bibfield  {title} {\bibinfo {title} {Strong
  and weak chaos in nonlinear networks with time-delayed couplings},\
  }\href@noop {} {\bibfield  {journal} {\bibinfo  {journal} {Physical Review
  Letters}\ }\textbf {\bibinfo {volume} {107}},\ \bibinfo {pages} {234102}
  (\bibinfo {year} {2011})}\BibitemShut {NoStop}%
\bibitem [{\citenamefont {Nair}\ \emph {et~al.}(2021)\citenamefont {Nair},
  \citenamefont {Hu}, \citenamefont {Berrill}, \citenamefont {Wiesenfeld},\
  and\ \citenamefont {Braiman}}]{nair2021using}%
  \BibitemOpen
  \bibfield  {author} {\bibinfo {author} {\bibfnamefont {N.}~\bibnamefont
  {Nair}}, \bibinfo {author} {\bibfnamefont {K.}~\bibnamefont {Hu}}, \bibinfo
  {author} {\bibfnamefont {M.}~\bibnamefont {Berrill}}, \bibinfo {author}
  {\bibfnamefont {K.}~\bibnamefont {Wiesenfeld}},\ and\ \bibinfo {author}
  {\bibfnamefont {Y.}~\bibnamefont {Braiman}},\ }\bibfield  {title} {\bibinfo
  {title} {Using disorder to overcome disorder: A mechanism for frequency and
  phase synchronization of diode laser arrays},\ }\href@noop {} {\bibfield
  {journal} {\bibinfo  {journal} {Physical Review Letters}\ }\textbf {\bibinfo
  {volume} {127}},\ \bibinfo {pages} {173901} (\bibinfo {year}
  {2021})}\BibitemShut {NoStop}%
\bibitem [{\citenamefont {Barioni}\ \emph {et~al.}(2025)\citenamefont
  {Barioni}, \citenamefont {Montanari},\ and\ \citenamefont
  {Motter}}]{barioni2025interpretable}%
  \BibitemOpen
  \bibfield  {author} {\bibinfo {author} {\bibfnamefont {A.~E.~D.}\
  \bibnamefont {Barioni}}, \bibinfo {author} {\bibfnamefont {A.~N.}\
  \bibnamefont {Montanari}},\ and\ \bibinfo {author} {\bibfnamefont {A.~E.}\
  \bibnamefont {Motter}},\ }\bibfield  {title} {\bibinfo {title} {Interpretable
  disorder-promoted synchronization and coherence in coupled laser networks},\
  }\href@noop {} {\bibfield  {journal} {\bibinfo  {journal} {Physical Review
  Letters}\ }\textbf {\bibinfo {volume} {135}},\ \bibinfo {pages} {197401}
  (\bibinfo {year} {2025})}\BibitemShut {NoStop}%
\bibitem [{\citenamefont {Ott}\ and\ \citenamefont
  {Antonsen}(2008)}]{ott2008low}%
  \BibitemOpen
  \bibfield  {author} {\bibinfo {author} {\bibfnamefont {E.}~\bibnamefont
  {Ott}}\ and\ \bibinfo {author} {\bibfnamefont {T.~M.}\ \bibnamefont
  {Antonsen}},\ }\bibfield  {title} {\bibinfo {title} {Low dimensional behavior
  of large systems of globally coupled oscillators},\ }\href@noop {} {\bibfield
   {journal} {\bibinfo  {journal} {Chaos: An Interdisciplinary Journal of
  Nonlinear Science}\ }\textbf {\bibinfo {volume} {18}},\ \bibinfo {pages}
  {037113} (\bibinfo {year} {2008})}\BibitemShut {NoStop}%
\bibitem [{\citenamefont {Rosenblum}\ and\ \citenamefont
  {Pikovsky}(2019)}]{rosenblum2019numerical}%
  \BibitemOpen
  \bibfield  {author} {\bibinfo {author} {\bibfnamefont {M.}~\bibnamefont
  {Rosenblum}}\ and\ \bibinfo {author} {\bibfnamefont {A.}~\bibnamefont
  {Pikovsky}},\ }\bibfield  {title} {\bibinfo {title} {Numerical phase
  reduction beyond the first order approximation},\ }\href@noop {} {\bibfield
  {journal} {\bibinfo  {journal} {Chaos: An Interdisciplinary Journal of
  Nonlinear Science}\ }\textbf {\bibinfo {volume} {29}},\ \bibinfo {pages}
  {011105} (\bibinfo {year} {2019})}\BibitemShut {NoStop}%
\bibitem [{\citenamefont {Mau}\ \emph {et~al.}(2023)\citenamefont {Mau},
  \citenamefont {Rosenblum},\ and\ \citenamefont {Pikovsky}}]{mau2023high}%
  \BibitemOpen
  \bibfield  {author} {\bibinfo {author} {\bibfnamefont {E.~T.}\ \bibnamefont
  {Mau}}, \bibinfo {author} {\bibfnamefont {M.}~\bibnamefont {Rosenblum}},\
  and\ \bibinfo {author} {\bibfnamefont {A.}~\bibnamefont {Pikovsky}},\
  }\bibfield  {title} {\bibinfo {title} {High-order phase reduction for coupled
  2d oscillators},\ }\href@noop {} {\bibfield  {journal} {\bibinfo  {journal}
  {Chaos: an Interdisciplinary Journal of Nonlinear Science}\ }\textbf
  {\bibinfo {volume} {33}},\ \bibinfo {pages} {10101} (\bibinfo {year}
  {2023})}\BibitemShut {NoStop}%
\bibitem [{\citenamefont {Gengel}\ \emph {et~al.}(2020)\citenamefont {Gengel},
  \citenamefont {Teichmann}, \citenamefont {Rosenblum},\ and\ \citenamefont
  {Pikovsky}}]{gengel2020high}%
  \BibitemOpen
  \bibfield  {author} {\bibinfo {author} {\bibfnamefont {E.}~\bibnamefont
  {Gengel}}, \bibinfo {author} {\bibfnamefont {E.}~\bibnamefont {Teichmann}},
  \bibinfo {author} {\bibfnamefont {M.}~\bibnamefont {Rosenblum}},\ and\
  \bibinfo {author} {\bibfnamefont {A.}~\bibnamefont {Pikovsky}},\ }\bibfield
  {title} {\bibinfo {title} {High-order phase reduction for coupled
  oscillators},\ }\href@noop {} {\bibfield  {journal} {\bibinfo  {journal}
  {Journal of Physics: Complexity}\ }\textbf {\bibinfo {volume} {2}},\ \bibinfo
  {pages} {015005} (\bibinfo {year} {2020})}\BibitemShut {NoStop}%
\bibitem [{\citenamefont {Mau}\ \emph {et~al.}(2024)\citenamefont {Mau},
  \citenamefont {Omelchenko},\ and\ \citenamefont {Rosenblum}}]{mau2024phase}%
  \BibitemOpen
  \bibfield  {author} {\bibinfo {author} {\bibfnamefont {E.~T.}\ \bibnamefont
  {Mau}}, \bibinfo {author} {\bibfnamefont {O.~E.}\ \bibnamefont
  {Omelchenko}},\ and\ \bibinfo {author} {\bibfnamefont {M.}~\bibnamefont
  {Rosenblum}},\ }\bibfield  {title} {\bibinfo {title} {Phase reduction
  explains chimera shape: When multibody interaction matters},\ }\href@noop {}
  {\bibfield  {journal} {\bibinfo  {journal} {Physical Review E}\ }\textbf
  {\bibinfo {volume} {110}},\ \bibinfo {pages} {L022201} (\bibinfo {year}
  {2024})}\BibitemShut {NoStop}%
\bibitem [{\citenamefont {Kotani}\ \emph {et~al.}(2020)\citenamefont {Kotani},
  \citenamefont {Ogawa}, \citenamefont {Shirasaka}, \citenamefont {Akao},
  \citenamefont {Jimbo},\ and\ \citenamefont {Nakao}}]{Kotani2020}%
  \BibitemOpen
  \bibfield  {author} {\bibinfo {author} {\bibfnamefont {K.}~\bibnamefont
  {Kotani}}, \bibinfo {author} {\bibfnamefont {Y.}~\bibnamefont {Ogawa}},
  \bibinfo {author} {\bibfnamefont {S.}~\bibnamefont {Shirasaka}}, \bibinfo
  {author} {\bibfnamefont {A.}~\bibnamefont {Akao}}, \bibinfo {author}
  {\bibfnamefont {Y.}~\bibnamefont {Jimbo}},\ and\ \bibinfo {author}
  {\bibfnamefont {H.}~\bibnamefont {Nakao}},\ }\bibfield  {title} {\bibinfo
  {title} {Nonlinear phase-amplitude reduction of delay-induced oscillations},\
  }\href@noop {} {\bibfield  {journal} {\bibinfo  {journal} {Phys. Rev.
  Research}\ }\textbf {\bibinfo {volume} {2}},\ \bibinfo {pages} {033106}
  (\bibinfo {year} {2020})}\BibitemShut {NoStop}%
\bibitem [{\citenamefont {Nicks}\ \emph {et~al.}(2024)\citenamefont {Nicks},
  \citenamefont {Allen},\ and\ \citenamefont {Coombes}}]{Nicks2024}%
  \BibitemOpen
  \bibfield  {author} {\bibinfo {author} {\bibfnamefont {R.}~\bibnamefont
  {Nicks}}, \bibinfo {author} {\bibfnamefont {R.}~\bibnamefont {Allen}},\ and\
  \bibinfo {author} {\bibfnamefont {S.}~\bibnamefont {Coombes}},\ }\bibfield
  {title} {\bibinfo {title} {Phase and amplitude responses for delay equations
  using harmonic balance},\ }\href@noop {} {\bibfield  {journal} {\bibinfo
  {journal} {Physical Review E}\ }\textbf {\bibinfo {volume} {110}},\ \bibinfo
  {pages} {L012202} (\bibinfo {year} {2024})}\BibitemShut {NoStop}%
\bibitem [{Note1()}]{Note1}%
  \BibitemOpen
  \bibinfo {note} {After submission of the present manuscript, a related
  preprint by N.~Fujii, K.~Taga, R.~Muolo, B.~Rink, and H.~Nakao
  (arXiv:2512.16193) appeared deriving a phase-only effective model with
  two-body and three-body interactions for globally coupled Kuramoto
  oscillators, with emphasis on synchronization transitions and Ott--Antonsen
  analysis. In the special case of sinusoidal coupling and weak frequency
  heterogeneity, Eq.~(S.12) of our Supplementary Material is closely related in
  structure to their reduced equations. In contrast, the present work develops
  a compact inertial second-order reduction for broad time-delayed
  Kuramoto--Daido networks with arbitrary topology, heterogeneous frequencies,
  and higher-harmonic coupling, aimed at predicting high-dimensional collective
  states and their statistics.}\BibitemShut {Stop}%
\bibitem [{\citenamefont {Acebr{\'o}n}\ \emph {et~al.}(2005)\citenamefont
  {Acebr{\'o}n}, \citenamefont {Bonilla}, \citenamefont {Vicente},
  \citenamefont {Ritort},\ and\ \citenamefont {Spigler}}]{acebron}%
  \BibitemOpen
  \bibfield  {author} {\bibinfo {author} {\bibfnamefont {J.~A.}\ \bibnamefont
  {Acebr{\'o}n}}, \bibinfo {author} {\bibfnamefont {L.~L.}\ \bibnamefont
  {Bonilla}}, \bibinfo {author} {\bibfnamefont {C.~J.~P.}\ \bibnamefont
  {Vicente}}, \bibinfo {author} {\bibfnamefont {F.}~\bibnamefont {Ritort}},\
  and\ \bibinfo {author} {\bibfnamefont {R.}~\bibnamefont {Spigler}},\
  }\bibfield  {title} {\bibinfo {title} {The {Kuramoto} model: A simple
  paradigm for synchronization phenomena},\ }\href@noop {} {\bibfield
  {journal} {\bibinfo  {journal} {Reviews of Modern Physics}\ }\textbf
  {\bibinfo {volume} {77}},\ \bibinfo {pages} {137} (\bibinfo {year}
  {2005})}\BibitemShut {NoStop}%
\bibitem [{\citenamefont {Smirnov}\ \emph {et~al.}(2017)\citenamefont
  {Smirnov}, \citenamefont {Osipov},\ and\ \citenamefont
  {Pikovsky}}]{smirnov2017chimera}%
  \BibitemOpen
  \bibfield  {author} {\bibinfo {author} {\bibfnamefont {L.}~\bibnamefont
  {Smirnov}}, \bibinfo {author} {\bibfnamefont {G.}~\bibnamefont {Osipov}},\
  and\ \bibinfo {author} {\bibfnamefont {A.}~\bibnamefont {Pikovsky}},\
  }\bibfield  {title} {\bibinfo {title} {Chimera patterns in the
  kuramoto--battogtokh model},\ }\href@noop {} {\bibfield  {journal} {\bibinfo
  {journal} {Journal of Physics A: Mathematical and Theoretical}\ }\textbf
  {\bibinfo {volume} {50}},\ \bibinfo {pages} {08LT01} (\bibinfo {year}
  {2017})}\BibitemShut {NoStop}%
\bibitem [{\citenamefont {Munyayev}\ \emph {et~al.}(2023)\citenamefont
  {Munyayev}, \citenamefont {Bolotov}, \citenamefont {Smirnov}, \citenamefont
  {Osipov},\ and\ \citenamefont {Belykh}}]{munyayev2023cyclops}%
  \BibitemOpen
  \bibfield  {author} {\bibinfo {author} {\bibfnamefont {V.~O.}\ \bibnamefont
  {Munyayev}}, \bibinfo {author} {\bibfnamefont {M.~I.}\ \bibnamefont
  {Bolotov}}, \bibinfo {author} {\bibfnamefont {L.~A.}\ \bibnamefont
  {Smirnov}}, \bibinfo {author} {\bibfnamefont {G.~V.}\ \bibnamefont
  {Osipov}},\ and\ \bibinfo {author} {\bibfnamefont {I.}~\bibnamefont
  {Belykh}},\ }\bibfield  {title} {\bibinfo {title} {Cyclops states in
  repulsive {K}uramoto networks: the role of higher-order coupling},\
  }\href@noop {} {\bibfield  {journal} {\bibinfo  {journal} {Physical Review
  Letters}\ }\textbf {\bibinfo {volume} {130}},\ \bibinfo {pages} {107201}
  (\bibinfo {year} {2023})}\BibitemShut {NoStop}%
\bibitem [{\citenamefont {Berner}\ \emph {et~al.}(2021)\citenamefont {Berner},
  \citenamefont {Yanchuk}, \citenamefont {Maistrenko},\ and\ \citenamefont
  {Scholl}}]{berner2021generalized}%
  \BibitemOpen
  \bibfield  {author} {\bibinfo {author} {\bibfnamefont {R.}~\bibnamefont
  {Berner}}, \bibinfo {author} {\bibfnamefont {S.}~\bibnamefont {Yanchuk}},
  \bibinfo {author} {\bibfnamefont {Y.}~\bibnamefont {Maistrenko}},\ and\
  \bibinfo {author} {\bibfnamefont {E.}~\bibnamefont {Scholl}},\ }\bibfield
  {title} {\bibinfo {title} {Generalized splay states in phase oscillator
  networks},\ }\href@noop {} {\bibfield  {journal} {\bibinfo  {journal} {Chaos:
  An Interdisciplinary Journal of Nonlinear Science}\ }\textbf {\bibinfo
  {volume} {31}},\ \bibinfo {pages} {073128} (\bibinfo {year}
  {2021})}\BibitemShut {NoStop}%
\bibitem [{\citenamefont {Munyayev}\ \emph {et~al.}(2022)\citenamefont
  {Munyayev}, \citenamefont {Bolotov}, \citenamefont {Smirnov}, \citenamefont
  {Osipov},\ and\ \citenamefont {Belykh}}]{munyayev2022stability}%
  \BibitemOpen
  \bibfield  {author} {\bibinfo {author} {\bibfnamefont {V.~O.}\ \bibnamefont
  {Munyayev}}, \bibinfo {author} {\bibfnamefont {M.~I.}\ \bibnamefont
  {Bolotov}}, \bibinfo {author} {\bibfnamefont {L.~A.}\ \bibnamefont
  {Smirnov}}, \bibinfo {author} {\bibfnamefont {G.~V.}\ \bibnamefont
  {Osipov}},\ and\ \bibinfo {author} {\bibfnamefont {I.~V.}\ \bibnamefont
  {Belykh}},\ }\bibfield  {title} {\bibinfo {title} {Stability of rotatory
  solitary states in kuramoto networks with inertia},\ }\href@noop {}
  {\bibfield  {journal} {\bibinfo  {journal} {Physical Review E}\ }\textbf
  {\bibinfo {volume} {105}},\ \bibinfo {pages} {024203} (\bibinfo {year}
  {2022})}\BibitemShut {NoStop}%
\bibitem [{\citenamefont {Earl}\ and\ \citenamefont
  {Strogatz}(2003{\natexlab{b}})}]{earl2003delay}%
  \BibitemOpen
  \bibfield  {author} {\bibinfo {author} {\bibfnamefont {M.~G.}\ \bibnamefont
  {Earl}}\ and\ \bibinfo {author} {\bibfnamefont {S.~H.}\ \bibnamefont
  {Strogatz}},\ }\bibfield  {title} {\bibinfo {title} {Synchronization in
  oscillator networks with delayed coupling: {A} stability criterion},\
  }\href@noop {} {\bibfield  {journal} {\bibinfo  {journal} {Physical Review
  E}\ }\textbf {\bibinfo {volume} {67}},\ \bibinfo {pages} {036204} (\bibinfo
  {year} {2003}{\natexlab{b}})}\BibitemShut {NoStop}%
\bibitem [{\citenamefont {Kuramoto}\ and\ \citenamefont
  {Battogtokh}(2002)}]{kuramoto2002coexistence}%
  \BibitemOpen
  \bibfield  {author} {\bibinfo {author} {\bibfnamefont {Y.}~\bibnamefont
  {Kuramoto}}\ and\ \bibinfo {author} {\bibfnamefont {D.}~\bibnamefont
  {Battogtokh}},\ }\bibfield  {title} {\bibinfo {title} {Coexistence of
  coherence and incoherence in nonlocally coupled phase oscillators},\
  }\href@noop {} {\bibfield  {journal} {\bibinfo  {journal} {Nonlinear
  Phenomena in Complex Systems}\ }\textbf {\bibinfo {volume} {5}},\ \bibinfo
  {pages} {380} (\bibinfo {year} {2002})}\BibitemShut {NoStop}%
\bibitem [{\citenamefont {Watts}\ and\ \citenamefont
  {Strogatz}(1998)}]{watts1998collective}%
  \BibitemOpen
  \bibfield  {author} {\bibinfo {author} {\bibfnamefont {D.~J.}\ \bibnamefont
  {Watts}}\ and\ \bibinfo {author} {\bibfnamefont {S.~H.}\ \bibnamefont
  {Strogatz}},\ }\bibfield  {title} {\bibinfo {title} {Collective dynamics of
  ‘small-world’networks},\ }\href@noop {} {\bibfield  {journal} {\bibinfo
  {journal} {Nature}\ }\textbf {\bibinfo {volume} {393}},\ \bibinfo {pages}
  {440} (\bibinfo {year} {1998})}\BibitemShut {NoStop}%
\bibitem [{\citenamefont {Bolotov}\ \emph {et~al.}(2024)\citenamefont
  {Bolotov}, \citenamefont {Munyayev}, \citenamefont {Smirnov}, \citenamefont
  {Osipov},\ and\ \citenamefont {Belykh}}]{bolotov2024breathing}%
  \BibitemOpen
  \bibfield  {author} {\bibinfo {author} {\bibfnamefont {M.~I.}\ \bibnamefont
  {Bolotov}}, \bibinfo {author} {\bibfnamefont {V.~O.}\ \bibnamefont
  {Munyayev}}, \bibinfo {author} {\bibfnamefont {L.~A.}\ \bibnamefont
  {Smirnov}}, \bibinfo {author} {\bibfnamefont {G.~V.}\ \bibnamefont
  {Osipov}},\ and\ \bibinfo {author} {\bibfnamefont {I.}~\bibnamefont
  {Belykh}},\ }\bibfield  {title} {\bibinfo {title} {Breathing and switching
  cyclops states in kuramoto networks with higher-mode coupling},\ }\href@noop
  {} {\bibfield  {journal} {\bibinfo  {journal} {Physical Review E}\ }\textbf
  {\bibinfo {volume} {109}},\ \bibinfo {pages} {054202} (\bibinfo {year}
  {2024})}\BibitemShut {NoStop}%
\bibitem [{\citenamefont {Bolotov}\ \emph {et~al.}(2025)\citenamefont
  {Bolotov}, \citenamefont {Smirnov}, \citenamefont {Munyayev}, \citenamefont
  {Osipov},\ and\ \citenamefont {Belykh}}]{bolotov2025heterogeneity}%
  \BibitemOpen
  \bibfield  {author} {\bibinfo {author} {\bibfnamefont {M.~I.}\ \bibnamefont
  {Bolotov}}, \bibinfo {author} {\bibfnamefont {L.~A.}\ \bibnamefont
  {Smirnov}}, \bibinfo {author} {\bibfnamefont {V.~O.}\ \bibnamefont
  {Munyayev}}, \bibinfo {author} {\bibfnamefont {G.~V.}\ \bibnamefont
  {Osipov}},\ and\ \bibinfo {author} {\bibfnamefont {I.}~\bibnamefont
  {Belykh}},\ }\bibfield  {title} {\bibinfo {title} {Heterogeneity induces
  cyclops states in kuramoto networks with higher-mode coupling},\ }\href@noop
  {} {\bibfield  {journal} {\bibinfo  {journal} {Physical Review E}\ }\textbf
  {\bibinfo {volume} {112}},\ \bibinfo {pages} {L052202} (\bibinfo {year}
  {2025})}\BibitemShut {NoStop}%
\bibitem [{\citenamefont {O’Keeffe}\ \emph {et~al.}(2017)\citenamefont
  {O’Keeffe}, \citenamefont {Hong},\ and\ \citenamefont
  {Strogatz}}]{Keeffe2017}%
  \BibitemOpen
  \bibfield  {author} {\bibinfo {author} {\bibfnamefont {K.~P.}\ \bibnamefont
  {O’Keeffe}}, \bibinfo {author} {\bibfnamefont {H.}~\bibnamefont {Hong}},\
  and\ \bibinfo {author} {\bibfnamefont {S.~H.}\ \bibnamefont {Strogatz}},\
  }\bibfield  {title} {\bibinfo {title} {Oscillators that sync and swarm},\
  }\href@noop {} {\bibfield  {journal} {\bibinfo  {journal} {Nature
  communications}\ }\textbf {\bibinfo {volume} {8}},\ \bibinfo {pages} {1504}
  (\bibinfo {year} {2017})}\BibitemShut {NoStop}%
\bibitem [{\citenamefont {Yoon}\ \emph {et~al.}(2022)\citenamefont {Yoon},
  \citenamefont {O'Keeffe}, \citenamefont {Mendes},\ and\ \citenamefont
  {Goltsev}}]{Yoon2022}%
  \BibitemOpen
  \bibfield  {author} {\bibinfo {author} {\bibfnamefont {S.}~\bibnamefont
  {Yoon}}, \bibinfo {author} {\bibfnamefont {K.~P.}\ \bibnamefont {O'Keeffe}},
  \bibinfo {author} {\bibfnamefont {J.~F.~F.}\ \bibnamefont {Mendes}},\ and\
  \bibinfo {author} {\bibfnamefont {A.~V.}\ \bibnamefont {Goltsev}},\
  }\bibfield  {title} {\bibinfo {title} {Sync and swarm: Solvable model of
  nonidentical swarmalators},\ }\href
  {https://doi.org/10.1103/PhysRevLett.129.208002} {\bibfield  {journal}
  {\bibinfo  {journal} {Phys. Rev. Lett.}\ }\textbf {\bibinfo {volume} {129}},\
  \bibinfo {pages} {208002} (\bibinfo {year} {2022})}\BibitemShut {NoStop}%
\bibitem [{\citenamefont {Hong}\ \emph {et~al.}(2023)\citenamefont {Hong},
  \citenamefont {O'Keeffe}, \citenamefont {Lee},\ and\ \citenamefont
  {Park}}]{Hong2023}%
  \BibitemOpen
  \bibfield  {author} {\bibinfo {author} {\bibfnamefont {H.}~\bibnamefont
  {Hong}}, \bibinfo {author} {\bibfnamefont {K.~P.}\ \bibnamefont {O'Keeffe}},
  \bibinfo {author} {\bibfnamefont {J.~S.}\ \bibnamefont {Lee}},\ and\ \bibinfo
  {author} {\bibfnamefont {H.}~\bibnamefont {Park}},\ }\bibfield  {title}
  {\bibinfo {title} {Swarmalators with thermal noise},\ }\href
  {https://doi.org/10.1103/PhysRevResearch.5.023105} {\bibfield  {journal}
  {\bibinfo  {journal} {Phys. Rev. Res.}\ }\textbf {\bibinfo {volume} {5}},\
  \bibinfo {pages} {023105} (\bibinfo {year} {2023})}\BibitemShut {NoStop}%
\bibitem [{\citenamefont {Liz\'arraga}\ and\ \citenamefont
  {de~Aguiar}(2023)}]{Lizarraga2023}%
  \BibitemOpen
  \bibfield  {author} {\bibinfo {author} {\bibfnamefont {J.~U.~F.}\
  \bibnamefont {Liz\'arraga}}\ and\ \bibinfo {author} {\bibfnamefont
  {M.~A.~M.}\ \bibnamefont {de~Aguiar}},\ }\bibfield  {title} {\bibinfo {title}
  {Synchronization of sakaguchi swarmalators},\ }\href
  {https://doi.org/10.1103/PhysRevE.108.024212} {\bibfield  {journal} {\bibinfo
   {journal} {Phys. Rev. E}\ }\textbf {\bibinfo {volume} {108}},\ \bibinfo
  {pages} {024212} (\bibinfo {year} {2023})}\BibitemShut {NoStop}%
\bibitem [{\citenamefont {Cai}\ \emph {et~al.}(2024)\citenamefont {Cai},
  \citenamefont {Liu}, \citenamefont {Guan}, \citenamefont {Kurths},\ and\
  \citenamefont {Zou}}]{Cai2024}%
  \BibitemOpen
  \bibfield  {author} {\bibinfo {author} {\bibfnamefont {Z.}~\bibnamefont
  {Cai}}, \bibinfo {author} {\bibfnamefont {Z.}~\bibnamefont {Liu}}, \bibinfo
  {author} {\bibfnamefont {S.}~\bibnamefont {Guan}}, \bibinfo {author}
  {\bibfnamefont {J.}~\bibnamefont {Kurths}},\ and\ \bibinfo {author}
  {\bibfnamefont {Y.}~\bibnamefont {Zou}},\ }\bibfield  {title} {\bibinfo
  {title} {High-mode coupling yields multicoherent-phase phenomena in
  nonlocally coupled oscillators},\ }\href
  {https://doi.org/10.1103/PhysRevLett.133.227201} {\bibfield  {journal}
  {\bibinfo  {journal} {Phys. Rev. Lett.}\ }\textbf {\bibinfo {volume} {133}},\
  \bibinfo {pages} {227201} (\bibinfo {year} {2024})}\BibitemShut {NoStop}%
\bibitem [{\citenamefont {O'Keeffe}\ and\ \citenamefont
  {Hindes}(2026)}]{Keeffe2026}%
  \BibitemOpen
  \bibfield  {author} {\bibinfo {author} {\bibfnamefont {K.}~\bibnamefont
  {O'Keeffe}}\ and\ \bibinfo {author} {\bibfnamefont {J.}~\bibnamefont
  {Hindes}},\ }\bibfield  {title} {\bibinfo {title} {Time delay in the 1d
  swarmalator model},\ }\href@noop {} {\bibfield  {journal} {\bibinfo
  {journal} {arXiv:2602.08156}\ } (\bibinfo {year} {2026})}\BibitemShut
  {NoStop}%
\bibitem [{\citenamefont {Ermentrout}(1991)}]{ermentrout1991adaptive}%
  \BibitemOpen
  \bibfield  {author} {\bibinfo {author} {\bibfnamefont {B.}~\bibnamefont
  {Ermentrout}},\ }\bibfield  {title} {\bibinfo {title} {An adaptive model for
  synchrony in the firefly pteroptyx malaccae},\ }\href@noop {} {\bibfield
  {journal} {\bibinfo  {journal} {Journal of Mathematical Biology}\ }\textbf
  {\bibinfo {volume} {29}},\ \bibinfo {pages} {571} (\bibinfo {year}
  {1991})}\BibitemShut {NoStop}%
\bibitem [{\citenamefont {Tumash}\ \emph {et~al.}(2019)\citenamefont {Tumash},
  \citenamefont {Olmi},\ and\ \citenamefont
  {Sch{\"o}ll}}]{tumash2019stability}%
  \BibitemOpen
  \bibfield  {author} {\bibinfo {author} {\bibfnamefont {L.}~\bibnamefont
  {Tumash}}, \bibinfo {author} {\bibfnamefont {S.}~\bibnamefont {Olmi}},\ and\
  \bibinfo {author} {\bibfnamefont {E.}~\bibnamefont {Sch{\"o}ll}},\ }\bibfield
   {title} {\bibinfo {title} {Stability and control of power grids with diluted
  network topology},\ }\href@noop {} {\bibfield  {journal} {\bibinfo  {journal}
  {Chaos: An Interdisciplinary Journal of Nonlinear Science}\ }\textbf
  {\bibinfo {volume} {29}},\ \bibinfo {pages} {123105} (\bibinfo {year}
  {2019})}\BibitemShut {NoStop}%
\bibitem [{\citenamefont {Gerstner}\ and\ \citenamefont
  {Kistler}(2002)}]{gerstner2002spiking}%
  \BibitemOpen
  \bibfield  {author} {\bibinfo {author} {\bibfnamefont {W.}~\bibnamefont
  {Gerstner}}\ and\ \bibinfo {author} {\bibfnamefont {W.~M.}\ \bibnamefont
  {Kistler}},\ }\href@noop {} {\emph {\bibinfo {title} {Spiking Neuron Models:
  Single Neurons, Populations, Plasticity}}}\ (\bibinfo  {publisher} {Cambridge
  University Press},\ \bibinfo {year} {2002})\BibitemShut {NoStop}%
\bibitem [{\citenamefont {Wiesenfeld}\ \emph {et~al.}(1996)\citenamefont
  {Wiesenfeld}, \citenamefont {Colet},\ and\ \citenamefont
  {Strogatz}}]{wiesenfeld1996synchronization}%
  \BibitemOpen
  \bibfield  {author} {\bibinfo {author} {\bibfnamefont {K.}~\bibnamefont
  {Wiesenfeld}}, \bibinfo {author} {\bibfnamefont {P.}~\bibnamefont {Colet}},\
  and\ \bibinfo {author} {\bibfnamefont {S.~H.}\ \bibnamefont {Strogatz}},\
  }\bibfield  {title} {\bibinfo {title} {Synchronization transitions in a
  disordered josephson series array},\ }\href@noop {} {\bibfield  {journal}
  {\bibinfo  {journal} {Physical Review Letters}\ }\textbf {\bibinfo {volume}
  {76}},\ \bibinfo {pages} {404} (\bibinfo {year} {1996})}\BibitemShut
  {NoStop}%
\bibitem [{\citenamefont {Wiesenfeld}\ \emph {et~al.}(1998)\citenamefont
  {Wiesenfeld}, \citenamefont {Colet},\ and\ \citenamefont
  {Strogatz}}]{wiesenfeld1998frequency}%
  \BibitemOpen
  \bibfield  {author} {\bibinfo {author} {\bibfnamefont {K.}~\bibnamefont
  {Wiesenfeld}}, \bibinfo {author} {\bibfnamefont {P.}~\bibnamefont {Colet}},\
  and\ \bibinfo {author} {\bibfnamefont {S.~H.}\ \bibnamefont {Strogatz}},\
  }\bibfield  {title} {\bibinfo {title} {Frequency locking in josephson arrays:
  Connection with the kuramoto model},\ }\href@noop {} {\bibfield  {journal}
  {\bibinfo  {journal} {Physical Review E}\ }\textbf {\bibinfo {volume} {57}},\
  \bibinfo {pages} {1563} (\bibinfo {year} {1998})}\BibitemShut {NoStop}%
\bibitem [{\citenamefont {Trees}\ \emph {et~al.}(2005)\citenamefont {Trees},
  \citenamefont {Saranathan},\ and\ \citenamefont
  {Stroud}}]{trees2005synchronization}%
  \BibitemOpen
  \bibfield  {author} {\bibinfo {author} {\bibfnamefont {B.~R.}\ \bibnamefont
  {Trees}}, \bibinfo {author} {\bibfnamefont {V.}~\bibnamefont {Saranathan}},\
  and\ \bibinfo {author} {\bibfnamefont {D.}~\bibnamefont {Stroud}},\
  }\bibfield  {title} {\bibinfo {title} {Synchronization in disordered
  josephson junction arrays: Small-world connections and the kuramoto model},\
  }\href@noop {} {\bibfield  {journal} {\bibinfo  {journal} {Physical Review
  E}\ }\textbf {\bibinfo {volume} {71}},\ \bibinfo {pages} {016215} (\bibinfo
  {year} {2005})}\BibitemShut {NoStop}%
\bibitem [{\citenamefont {Garbin}\ \emph {et~al.}(2015)\citenamefont {Garbin},
  \citenamefont {Javaloyes}, \citenamefont {Tissoni},\ and\ \citenamefont
  {Barland}}]{garbin2015topological}%
  \BibitemOpen
  \bibfield  {author} {\bibinfo {author} {\bibfnamefont {B.}~\bibnamefont
  {Garbin}}, \bibinfo {author} {\bibfnamefont {J.}~\bibnamefont {Javaloyes}},
  \bibinfo {author} {\bibfnamefont {G.}~\bibnamefont {Tissoni}},\ and\ \bibinfo
  {author} {\bibfnamefont {S.}~\bibnamefont {Barland}},\ }\bibfield  {title}
  {\bibinfo {title} {{Topological solitons as addressable phase bits in a
  driven laser}},\ }\href@noop {} {\bibfield  {journal} {\bibinfo  {journal}
  {Nature Communications}\ }\textbf {\bibinfo {volume} {6}},\ \bibinfo {pages}
  {5915} (\bibinfo {year} {2015})}\BibitemShut {NoStop}%
\end{thebibliography}%

\clearpage
\onecolumngrid
\renewcommand{\figurename}{Supplementary Figure}
%\makeatother
%\setcounter{equation}{8}
\setcounter{equation}{0}
\renewcommand\theequation{S.\arabic{equation}}
\setcounter{figure}{0}
\section{Derivation of the Second-Order Reduced Model}
We study a generalized KD network with heterogeneous natural frequencies, external forcing, and time-delayed pairwise coupling:
\begin{equation}
	\dfrac{d\theta_{j}(t)}{dt} = \varpi + \eta_{j}(t )+ \frac{\varkappa}{N}\sum_{k=1}^{N}F_{jk}\bigl(\theta_{k}(t-\tau)-\theta_{j}(t)\bigr),
	\label{eq:kd-model}
\end{equation}
where $\theta_{j}(t)$ is the phase of oscillator $j$ ($j=1,\dots,N$), $\varpi$ is a baseline (mean) natural frequency, $\eta_j(t)$ collects deviations from this baseline (including external forcing or detuning), $\varkappa$ is the coupling strength, and $F_{jk}(\cdot)$ are $2\pi$-periodic pairwise coupling functions that may differ across oscillator pairs and represent, for example, random interactions. The parameter $\tau>0$ is a uniform coupling delay. 

\subsection{Basic Assumptions for Reduction: Weak Heterogeneity and Small Coupling Strength}
\label{subsec:assumptions}

To derive a reduced description of the dynamics governed by Eq.~\eqref{eq:kd-model}, we consider the regime in which both the heterogeneity $(\eta_j(t)$ and the coupling strength $\varkappa$ are weak. We introduce a small parameter $\varepsilon \ll 1$ and scale these quantities as
\[
\eta_{j}(t)=\varepsilon\omega_{j}+\varepsilon\zeta_{j}(t), \qquad 
\varkappa=\varepsilon\kappa.
\]
Here, $\omega_j$ denotes the (scaled) deviations of the natural frequencies from the mean value $\varpi$, $\zeta_j(t)$ represents additional time-dependent perturbations, and $\kappa$ is the unscaled coupling constant.

This setting corresponds to a self-oscillatory regime in which the dominant behavior is nearly uniform rotation at frequency $\varpi$. Indeed, in the limit $\varepsilon=0$, all oscillators evolve independently with $\dot{\theta}_j=\varpi$, yielding identical uniform rotation. The introduction of $\varepsilon$ thus enables a systematic perturbative treatment of weak heterogeneity and weak coupling around this uniform rotational state.

\subsection{Multiple Time Scale Expansion}
\label{subsec:mts-expansion}

The assumption that the oscillators exhibit nearly uniform rotation at the dominant frequency $\varpi$, with slow modulations induced by weak disorder, external influences, and pairwise interactions, enables the use of a multiple-timescale expansion. This perturbative technique is well-suited to systems with dynamics that evolve on distinct temporal scales.

To systematically capture the slow evolution of the phases, we introduce a hierarchy of time scales $t_s = \varepsilon^{s} t$ for $s=0,1,2,\ldots$. This separation allows us to distinguish the fast oscillatory motion from the slow modulations driven by heterogeneity and coupling. Before proceeding with the reduction, we make one technical remark that clarifies the structure of the analysis.

Because the variables $t_s=\varepsilon^{s}t$ are treated as independent, the total time derivative expands as a sum of partial derivatives:
\begin{equation}
	\frac{d}{dt}
	= \sum_{s=0}^{\infty} \varepsilon^{s} \frac{\partial}{\partial t_s}
	= \frac{\partial}{\partial t_0}
	+ \varepsilon \frac{\partial}{\partial t_1}
	+ \varepsilon^{2} \frac{\partial}{\partial t_2}
	+ \ldots .
	\label{eq:time-derivative}
\end{equation}

We seek solutions of Eq.~\eqref{eq:kd-model} in the form of an asymptotic expansion in powers of $\varepsilon$:
\begin{equation}
	\theta_{j}(t)
	= \varpi t_{0}
	+ \phi_{j}(t_{1},t_{2},\ldots)
	+ \sum_{p=1}^{\infty}
	\varepsilon^{p}
	\varphi_{j}^{(p)}(t_{0},t_{1},t_{2},\ldots),
	\label{eq:series}
\end{equation}
where the leading term $\varpi t_{0}$ represents the primary fast rotation, while $\phi_{j}(t_{1},t_{2},\ldots)$ describes the slow collective phase drift responsible for the emergent network-level dynamics. The functions $\varphi_{j}^{(p)}$ represent higher-order corrections that include rapid oscillatory components and transient effects, which average out over long time scales.

\subsection{Expansion of Delayed Phase Differences}
\label{subsec:ph-dif-expansion}

To substitute the multi–scale expansion into the Kuramoto--Daido model, we first evaluate the delayed phase difference $\theta_{k}(t-\tau)-\theta_{j}(t)$. Using the asymptotic series~\eqref{eq:series}, and assuming that the slow time variables $t_{1}, t_{2}, \ldots$ are only weakly affected by the finite delay $\tau$ at leading order (i.e., $\tau$ does not generate additional fast time scales), we obtain the expansion
\begin{gather}
	\theta_{k}(t-\tau)-\theta_{j}(t)
	= -\varpi \tau
	+ \phi_{k}(t_{1},t_{2},\ldots)
	- \phi_{j}(t_{1},t_{2},\ldots)
	\nonumber\\{} 
	- \varepsilon \tau \frac{\partial \phi_{k}}{\partial t_{1}}
	+ \varepsilon \varphi_{k}^{(1)}(t_{0}-\tau,t_{1},t_{2},\ldots)
	- \varepsilon \varphi_{j}^{(1)}(t_{0},t_{1},t_{2},\ldots)
	+ o(\varepsilon).
	\label{eq:diff-theta-expansion}
\end{gather}
Here, the term $-\varpi\tau$ reflects the difference in the rapid rotation between times $t$ and $t-\tau$. Since $\varpi$ is not assumed to satisfy any resonance conditions with $\tau$, this constant phase shift may take arbitrary values.
The next terms,
$
\phi_{k}(t_{1},t_{2},\ldots)-\phi_{j}(t_{1},t_{2},\ldots),
$
represent the slow phase difference between oscillators $k$ and $j$. The contribution $-\varepsilon \tau \, \partial \phi_{k}/\partial t_{1}$ accounts for the slow drift of oscillator $k$ over the delay interval $\tau$ when observed on the $t_{1}$ time scale. Although the variables $\phi_{j}$ evolve slowly compared with the fast rotation, their change over time $\tau$ must still be retained at order $\varepsilon$. The remaining terms in~\eqref{eq:diff-theta-expansion} contain the first-order fast corrections $\varphi^{(1)}_{j}$, evaluated at $t_{0}$ and $t_{0}-\tau$, respectively. These describe rapid fluctuations around the slow phase manifold. The $o(\varepsilon)$ term indicates truncation at first order, appropriate for small $\varepsilon$.

\subsection{Hierarchy of Equations by Order of Smallness}
\label{subsec:eqs-hierarchy}

Substituting the expansion~\eqref{eq:diff-theta-expansion} into Eq.~\eqref{eq:kd-model} and using the multi–scale representation of the time derivative, we obtain
\begin{equation*}
	\begin{gathered}
		\varepsilon\left(\pd{\phi_{j}\left(t_{1},t_{2},\ldots\right)}{t_{1}} +  \pd{\varphi_{j}^{(1)}\left(t_{0},t_{1},t_{2},\ldots\right)}{t_{0}}\right)+ \varepsilon^{2}\left(\pd{\phi_{j}\left(t_{1},t_{2},\ldots\right)}{t_{2}} +\pd{\varphi_{j}^{(1)}\left(t_{0},t_{1},t_{2},\ldots\right)}{t_{1}} + \pd{\varphi_{j}^{(2)}\left(t_{0},t_{1},t_{2},\ldots\right)}{t_{0}}\right) = {}\\{}
		\varepsilon\omega_{j} + \varepsilon\zeta_{j}(t_{0},t_{1},t_{2},\ldots) + \frac{\varepsilon\kappa}{N}\sum_{k=1}^{N}F_{jk}\left(\phi_{k}\left(t_{1},t_{2},\ldots\right)-\phi_{j}\left(t_{1},t_{2},\ldots\right)-\varpi\tau\right) {}\\{} 
		- \frac{\varepsilon^{2}\tau\kappa}{N}\sum_{k=1}^{N}F'_{jk}\left(\phi_{k}\left(t_{1},t_{2},\ldots\right)-\phi_{j}\left(t_{1},t_{2},\ldots\right)-\varpi\tau\right)\pd{\phi_{k}\left(t_{1},t_{2},\ldots\right)}{t_{1}} {}\\{}
		+ \frac{\varepsilon^{2}\kappa}{N}\sum_{k=1}^{N}F'_{jk}\bigl(\phi_{k}\left(t_{1},t_{2},\ldots\right)-\phi_{j}\left(t_{1},t_{2},\ldots\right)-\varpi\tau\bigr)\Bigl(\varphi_{k}^{(1)}\left(t_{0}\!-\!\tau,t_{1},t_{2},\ldots\right)-\varphi_{j}^{(1)}\left(t_{0},t_{1},t_{2},\ldots\right)\Bigr) + o\left(\varepsilon^{2}\right),
	\end{gathered}
\end{equation*}
where we have grouped terms according to powers of $\varepsilon$.
The contributions on the left-hand side arise from the multi-scale expansion of the time derivative and the asymptotic series for $\theta_{j}(t)$, while the right-hand side results from the scaled heterogeneity, $\omega_{j}$ and $\zeta_{j}(t)$, the scaled coupling constant $\kappa$, and the Taylor expansion of the coupling functions $F_{jk}$ around their arguments
$\phi_{k}-\phi_{j}-\varpi\tau$.

This substitution produces a hierarchy of equations, each corresponding to a particular power of $\varepsilon$.  
The solvability condition at each order (specifically, the removal of secular terms that would otherwise lead to unbounded growth in the fast time variable $t_{0}$) determines the evolution equations for the slow phase variables $\phi_{j}$.  
This procedure is analogous to the averaging method: secular terms encode resonant forcing on the slow manifold, and their elimination yields the correct slow dynamics governing long-term behavior.

In what follows, we analyze this hierarchy order by order, extracting the dynamical equations for the slow modulation of phases that ultimately produce the first- and second-order reduced models.

\subsection{Order $\varepsilon^{1}$: Leading Slow Dynamics and First Fast Correction}
For the Kuramoto--Daido model~\eqref{eq:kd-model}, the multiple--time--scale expansion yields, at order $\varepsilon^{1}$, an equation determining the first fast correction 
$\varphi_{j}^{(1)}(t_{0},t_{1},t_{2},\ldots)$.  
Collecting all terms proportional to $\varepsilon$ gives
\begin{equation}
	\pd{\varphi_{j}^{(1)}\left(t_{0},t_{1},t_{2},\ldots\right)}{t_{0}} = \zeta_{j}\left(t_{0},t_{1},t_{2},\ldots\right) - \bar{\zeta}_{j}\left(t_{1},t_{2},\ldots\right), \quad
	\bar{\zeta}_{j}\left(t_{1},t_{2},\ldots\right) = \dfrac{\varpi}{2\pi}\!\!\int\limits_{t-\pi\left/\varpi\right.}^{t+\pi\left/\varpi\right.}\!\!\zeta_{j}\left(\varsigma\right)d\varsigma,
	\label{eq:eq-varphi-1}
\end{equation}
Here, $\bar{\zeta}_{j}$ is the average of the external forcing over one fast rotation period $2\pi/\varpi$, so the right–hand side represents only the oscillatory (zero-mean) part of $\zeta_{j}$.  
Enforcing this equation ensures that no secular terms appear in $\varphi_{j}^{(1)}$, preventing unbounded growth in the fast time variable $t_{0}$.

The solvability condition at this order then yields the evolution equation for the slow phase $\phi_{j}(t_{1},t_{2},\ldots)$ on the $t_{1}$ time scale:
\begin{equation}
	\pd{\phi_{j}\left(t_{1},t_{2},\ldots\right)}{t_{1}} = \omega_{j} + \bar{\zeta}_{j}\left(t_{1},t_{2},\ldots\right) +\frac{\kappa}{N}\sum_{k=1}^{N}F_{jk}\bigl(\phi_{k}\left(t_{1},t_{2},\ldots\right)-\phi_{j}\left(t_{1},t_{2},\ldots\right)-\varpi\tau\bigr).
	\label{eq:eq-phi-t1}
\end{equation}
This is the first-order reduced Kuramoto--Daido equation: the delay appears only as an effective phase shift $\varpi\tau$, and the natural frequencies are modified by the averaged external perturbations $\bar{\zeta}_{j}$.

\noindent\textit{Remark (First-order reduction in the original time variable).}
If one stops at order $\varepsilon$ and rewrites the dynamics using the original time derivative $d/dt$ from Eq.~\eqref{eq:time-derivative}, then the reduced system takes the standard Kuramoto--Daido form
\begin{equation}
	\frac{d\phi_{j}}{dt} = \bar{\eta}_{j}(t)+ \frac{\varkappa}{N}\sum_{k=1}^{N}{F_{jk}\left(\phi_{k}-\phi_{j}-\varpi\tau\right)}.
	\label{eq:model_reduced_v1}
\end{equation}
where
$\bar{\eta}_{j}(t)
= \varepsilon \omega_{j}
+ \varepsilon \bar{\zeta}_{j}(t).$
Thus, $\bar{\eta}_{j}(t)$ represents the combined effect of the small intrinsic frequency deviation of oscillator $j$ from $\varpi$ and the averaged external forcing acting on it.  Terms of order $o(\varepsilon)$ are consistently omitted.

\subsection{Order $\varepsilon^{2}$: Second Fast Correction and Higher-Order Slow Dynamics}
At order $\varepsilon^{2}$, collecting all terms proportional to $\varepsilon^{2}$ yields
\begin{gather}
	\pd{\varphi_{j}^{(2)}\left(t_{0},t_{1},t_{2},\ldots\right)}{t_{0}}=-\pd{\varphi_{j}^{(1)}\left(t_{0},t_{1},t_{2},\ldots\right)}{t_{1}}+\nonumber{}\\{}
	+\frac{\kappa}{N}\sum_{k=1}^{N}F'_{jk}\bigl(\phi_{k}\left(t_{1},t_{2},\ldots\right)-\phi_{j}\left(t_{1},t_{2},\ldots\right)-\varpi\tau\bigr)\Bigl(\varphi_{k}^{(1)}\left(t_{0}-\tau,t_{1},t_{2},\ldots\right)-\varphi_{j}^{(1)}\left(t_{0},t_{1},t_{2},\ldots\right)\Bigr).
	\label{eq:eq-varphi-2}
\end{gather}
which defines the second fast correction $\varphi_{j}^{(2)}$ while ensuring the absence of secular terms.

The first term on the right-hand side accounts for the slow modulation of the first-order fast correction $\varphi_{j}^{(1)}$.  
Eliminating secular terms at this order gives the next slow-time equation for $\phi_{j}$:
\begin{equation}
	\pd{\phi_{j}\left(t_{1},t_{2},\ldots\right)}{t_{2}} = -\frac{\tau\kappa}{N}\sum_{k=1}^{N}F'_{jk}\bigl(\phi_{k}\left(t_{1},t_{2},\ldots\right)-\phi_{j}\left(t_{1},t_{2},\ldots\right)-\varpi\tau\bigr)\pd{\phi_{k}\!\left(t_{1},t_{2},\ldots\right)}{t_{1}},
	\label{eq:eq-phi-t2}
\end{equation}
which represents the second-order correction to the slow phase drift induced jointly by delay and coupling.

\noindent\textit{Explicit form obtained by substituting the first-order slow dynamics.}  
Using Eq.~\eqref{eq:eq-phi-t1} inside Eq.~\eqref{eq:eq-phi-t2} yields
\begin{gather}
	\pd{\phi_{j}\left(t_{1},t_{2},\ldots\right)}{t_{2}} = -\frac{\tau\kappa}{N}\sum_{k=1}^{N}\left(\omega_{k}+\bar{\zeta}_{k}\left(t_{1},t_{2},\ldots\right)\right)\!
	F'_{jk}\bigl(\phi_{k}\left(t_{1},t_{2},\ldots\right)-\phi_{j}\left(t_{1},t_{2},\ldots\right)-\varpi\tau\bigr)- \nonumber\\
	-\frac{\tau\kappa^{2}}{N^{2}}\sum_{k=1}^{N}\sum_{\ell=1}^{N} F'_{jk}\bigl(\phi_{k}\left(t_{1},t_{2},\ldots\right)-\phi_{j}\left(t_{1},t_{2},\ldots\right)-\varpi\tau\bigr)
	F_{k\ell}\bigl(\phi_{\ell}\left(t_{1},t_{2},\ldots\right)-\phi_{k}\left(t_{1},t_{2},\ldots\right)-\varpi\tau\bigr).
	\label{eq:eq-phi-t2-}
\end{gather}

\noindent\textit{Returning to the original time variable.}  
The total time derivatives of $\phi_{j}$ are, up to $o(\varepsilon^{2})$,
\begin{equation}
	\frac{d\phi_{j}}{dt}
	= \varepsilon \frac{\partial \phi_{j}}{\partial t_{1}}
	+ \varepsilon^{2} \frac{\partial \phi_{j}}{\partial t_{2}}
	+ o(\varepsilon^{2}),
	\qquad
	\frac{d^{2}\phi_{j}}{dt^{2}}
	= \varepsilon^{2} \frac{\partial^{2} \phi_{j}}{\partial t_{1}^{2}}
	+ o(\varepsilon^{2}),
	\label{eq:phi-derivatives}
\end{equation}
and substituting Eqs.~\eqref{eq:eq-phi-t1} and~\eqref{eq:eq-phi-t2-} into~\eqref{eq:phi-derivatives} yields a closed second-order reduced model.

Carrying out this substitution produces, up to $o(\varepsilon^{2})$,
\begin{gather}
	\frac{d\phi_{j}}{dt} = \bar{\eta}_{j}(t)+ \frac{\varkappa}{N}\sum_{k=1}^{N}{F_{jk}\left(\phi_{k}-\phi_{j}-\varpi\tau\right)} \nonumber\\
	-\frac{\tau\varkappa}{N}\sum_{k=1}^{N}\bar{\eta}_{k}(t) F'_{jk}\bigl(\phi_{k}-\phi_{j}-\varpi\tau\bigr) -
	\frac{\tau\varkappa^{2}}{N^{2}}\sum_{k=1}^{N}\sum_{\ell=1}^{N} F'_{j\ell}\bigl(\phi_{\ell}-\phi_{j}-\varpi\tau\bigr)F_{\ell k}\bigl(\phi_{k}-\phi_{\ell}-\varpi\tau\bigr).
	\label{eq:model_reduced_v2}
\end{gather}
In the last term, the dummy indices $k$ and $\ell$ were interchanged for convenience.

This equation is a delay-free second-order phase model incorporating delay-induced inertial and higher-order interaction terms.  
Although exact, its explicit double-sum structure may be cumbersome for analytical work, especially in heterogeneous or random networks, thereby motivating the more compact formulation given in the main text.

\subsection{Auxiliary Relations and an Alternative Second-Order Reduced Model with Inertia and Multibody Interaction}
\label{subsec:reduced-model}

To obtain a more compact and analytically convenient second-order reduction, we make use of an auxiliary identity derived from the first-order slow-time equation~\eqref{eq:eq-phi-t1}.  
Differentiating Eq.~\eqref{eq:eq-phi-t1} with respect to $t_{1}$ gives
\begin{gather*}
	\pdd{\phi_{j}\left(t_{1},t_{2},\ldots\right)}{t_{1}} = \pd{\bar{\zeta}_{j}\left(t_{1},t_{2},\ldots\right)}{t_{1}} + \frac{\kappa}{N}\sum_{k=1}^{N}F'_{jk}\bigl(\phi_{k}\left(t_{1},t_{2},\ldots\right)-\phi_{j}\left(t_{1},t_{2},\ldots\bigr)-
	\varpi\tau\right)\pd{\phi_{k}\left(t_{1},t_{2},\ldots\right)}{t_{1}} \nonumber{}\\{}
	-\pd{\phi_{j}\left(t_{1},t_{2},\ldots\right)}{t_{1}}\frac{\kappa}{N}\sum_{k=1}^{N}F'_{jk}\!\bigl(\phi_{k}\left(t_{1},t_{2},\ldots\right)-\phi_{j}\left(t_{1},t_{2},\ldots\bigr)-\varpi\tau\right).
\end{gather*}
Using this identity, Eq.~\eqref{eq:eq-phi-t2} can be rewritten as
\begin{gather}
	\pd{\phi_{j}\left(t_{1},t_{2},\ldots\right)}{t_{2}} + \tau\pdd{\phi_{j}\left(t_{1},t_{2},\ldots\right)}{t_{1}} = \tau\pd{\bar{\zeta}_{j}\left(t_{1},t_{2},\ldots\right)}{t_{1}}\nonumber {}\\{}
	-\frac{\tau\kappa}{N}\pd{\phi_{j}\left(t_{1},t_{2},\ldots\right)}{t_{1}}\sum_{k=1}^{N}F'_{jk}\bigl(\phi_{k}\left(t_{1},t_{2},\ldots\right)-\phi_{j}\left(t_{1},t_{2},\ldots\right)-\varpi\tau\bigr),
	\label{eq:eq-phi-t2+}
\end{gather}
which makes explicit the coupling between the slow second derivative of $\phi_{j}$ and derivatives of the coupling and averaged external perturbations.  
This form is particularly convenient for constructing a delay-free reduced model of second-order accuracy.

\noindent\textit{Returning to the original time variable.}
Substituting relations~\eqref{eq:phi-derivatives} into Eq.~\eqref{eq:eq-phi-t2+}, and replacing $\partial\phi_{j}/\partial t_{1}$ via Eq.~\eqref{eq:eq-phi-t1}, yields a closed second-order model for the slow phase $\phi_{j}$:
\begin{equation}
	\tau\frac{d^{2}\phi_{j}}{dt^{2}} + \frac{d\phi_{j}}{dt} = \varepsilon\left(\omega_{j} + \bar{\zeta}_{j}(t)+ \frac{\kappa}{N}\sum_{k=1}^{N}{F_{jk}\left(\phi_{k}-\phi_{j}-\varpi\tau\right)}\right)
	\left(1 - \frac{\varepsilon\tau\kappa}{N}\sum_{k=1}^{N}{F'_{jk}\!\left(\phi_{k}-\phi_{j}-\varpi\tau\right)}\right) + \varepsilon\tau\frac{d\bar{\zeta}_{j}}{dt}
	\label{eq:model_reduced_v3}
\end{equation}
with all terms $o(\varepsilon^{2})$ omitted.  
This equation represents the dynamics of the slow phase $\phi_j\left(t_{1},t_{2},\ldots\right)$ up to second order in $\varepsilon$.
Notably, Eq.~\eqref{eq:model_reduced_v3} is a second-order differential equation, where, along with the phase shift $\varpi\tau$, which is incorporated into the pairwise interaction function, the second derivative term, scaled by $\tau$, arises due to the time delay in the original model.
The right-hand side of~Eq.~\eqref{eq:model_reduced_v3} effectively represents the driving force for the slow phase evolution, modulated by a factor that depends on the derivative of the coupling function, reflecting the influence of the delay on the effective interaction strength.

\medskip

\noindent\textit{Substitution of the compact notation for averaged disorder.}  
Using $\bar{\eta}_{j}(t)=\varepsilon\omega_{j}+\varepsilon\bar{\zeta}_{j}(t)$ and $\varkappa=\varepsilon\kappa$, Eq.~\eqref{eq:model_reduced_v3} becomes
\begin{equation}
	\tau\frac{d^{2}\phi_{j}}{dt^{2}} + \frac{d\phi_{j}}{dt} = \left(\bar{\eta}_{j}(t)+ \frac{\varkappa}{N}\sum_{k=1}^{N}{F_{jk}\left(\phi_{k}-\phi_{j}-\varpi\tau\right)}\right)\left(1 - \frac{\tau\varkappa}{N}\sum_{k=1}^{N}{F'_{jk}\!\left(\phi_{k}-\phi_{j}-\varpi\tau\right)}\right) + \tau\dfrac{d\bar{\eta}_{j}(t)}{dt}.
	\label{eq:model_reduced_v4}
\end{equation}
The quantity $\bar{\eta}_{j}(t)$ combines intrinsic frequency heterogeneity and the averaged external perturbation acting on oscillator $j$.

Equation~\eqref{eq:model_reduced_v4} is precisely the second-order reduced model presented as Eq.~(4) in the main text.  
It encapsulates the leading dynamical consequences of finite delay in the weak-coupling regime: an emergent inertial term, delay-induced renormalization of the effective coupling, and triadic interaction terms.  
This delay-free second-order phase description retains the essential structure of the underlying Kuramoto--Daido dynamics while offering a substantially more tractable framework, both analytically and numerically, than the original time-delayed system~\eqref{eq:kd-model}.  
As demonstrated in the main text, this reduced model accurately predicts complex collective dynamics and provides an efficient tool for studying high-dimensional patterns in time-delayed oscillator networks.
\begin{figure}[h]
	\includegraphics[width=0.45\columnwidth]{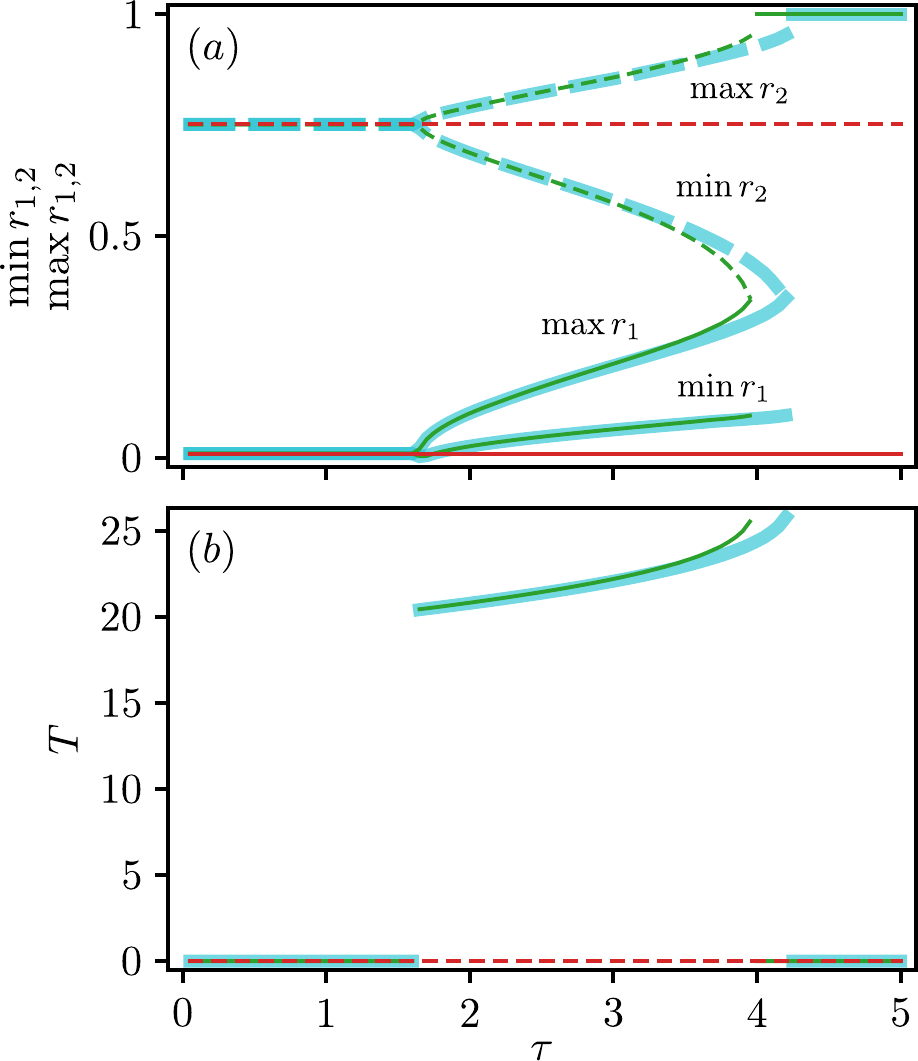}
	\caption{{Delay dependence of cyclops-state dynamics in the time-delayed biharmonic model~(1), with $F_{jk}$ as defined in the Global Bi-harmonic Coupling Section of the main text, and in its reduced descriptions. (a) Minimum and maximum values of the magnitudes of the first ($r_1$, solid) and second ($r_2$, dashed) Kuramoto order parameters versus $\tau$. (b) Period $T$ of the relative phase dynamics between the coherent clusters forming a cyclops state versus $\tau$. A nonzero value of $T$ identifies a breathing cyclops state, in which the inter-cluster phase differences evolve periodically in time. The phase shifts are chosen as $\alpha_1=1.62-1.187\tau$ and $\alpha_2=0.1-2.4\tau$, which keep the first-order reduced model unchanged and thereby isolate genuine second-order delay effects. Simulations are performed by continuation in $\tau$ with step $\Delta\tau=0.05$; for each new value of $\tau$, the final state from the previous run is used as the initial condition. Whenever $r_1\neq 0$, the realized regime is a cyclops state, either stationary or breathing. Colors: delayed model (cyan), first-order reduction (red), second-order reduction (green). The first-order reduction fails to predict the emergence of a nonstationary breathing cyclops state near $\tau\approx 1.6$ and instead yields an incorrect prediction of a stationary cyclops state for larger values of $\tau$. By contrast, the second-order reduction correctly predicts the breathing cyclops regime, characterized by distinct minimum and maximum values of $r_1$ and $r_2$, up to approximately $\tau\approx 3.5$, beyond which visible deviations from the benchmark delayed system begin to develop. Parameters: $N=9$, $\varkappa=0.1$, $K_1=1$, $K_2=0.05$, $\varpi=1.2$.}} 
	\label{fig:fig1}
\end{figure}
\section{Delay dependence and validity range of the second-order reduction}

{To probe the validity range of the second-order reduction, we consider the most demanding regime studied in the main text: nonstationary cyclops states in the globally coupled biharmonic network (1) (see Fig.~4). Here, we keep the coupling strength fixed and vary the delay $\tau$, so that the effective expansion parameter $\varkappa\tau$ increases systematically. At the same time, we choose the phase shifts $\alpha_1$ and $\alpha_2$ to depend linearly on $\tau$ in such a way that the corresponding first-order reduced model remains unchanged. This isolates the genuinely second-order effects of delay and provides a stringent test of how far the predictive power of the reduction extends beyond the formally asymptotic regime. As shown in Supplementary Fig.~\ref{fig:fig1}, the second-order model accurately tracks both the extrema of the order parameters and the breathing period up to approximately $\tau\approx 3.5$, i.e., $\varkappa\tau\approx 0.3.5$, whereas the first-order reduction fails to capture the nonstationary breathing dynamics.}

\section{Extension of the second-order delay reduction to time-delayed swarmalators}

\subsection{Time-delayed swarmalator model}

To further demonstrate the universality of the second-order reduction developed in the main text, we apply it here to a one-dimensional swarmalator model with time-delayed coupling in both spatial and phase dynamics. The reduction is carried out in the weak-coupling regime by a multiple-time-scale expansion, in direct analogy with the delayed Kuramoto--Daido case. As in the main text, the resulting delay-free description reveals two robust effects of delay: an effective inertial term generated by memory and delay-induced higher-order (triadic) interactions arising at second order.

Swarmalators are agents that self-organize simultaneously in physical space and in oscillatory phase \cite{Keeffe2017,Yoon2022,Hong2023,Lizarraga2023,Cai2024,Keeffe2026}. They provide a tractable framework for studying the interplay between swarming and synchronization. Here, we consider the following one-dimensional ring model of mobile oscillators:
\begin{gather}
	\frac{dx_{k}}{dt} = \nu_{k} + \frac{J}{N}\sum_{j=1}^{N} \sin\bigl(x_{j}(t-\tau)-x_{k}(t)-\beta\bigr)\cos\bigl(\theta_{j}(t-\tau)-\theta_{k}(t)-\alpha\bigr), \label{eq:del_x}\\
	\frac{d\theta_{k}}{dt} = \omega_{k} + \frac{K}{N}\sum_{j=1}^{N} \sin\bigl(\theta_{j}(t-\tau)-\theta_{k}(t)-\alpha\bigr)\cos\bigl(x_{j}(t-\tau)-x_{k}(t)-\beta\bigr). \label{eq:del_th}
\end{gather}

Recent studies \cite{Yoon2022,Hong2023,Lizarraga2023,Cai2024,Keeffe2026} 
have shown that uniform coupling delays in both spatial and phase interactions can generate unsteady collective regimes with time-periodic order parameters and bifurcations that can be analyzed exactly. At the same time, a systematic description of more complex behaviors, including chimera-like partially coherent states, remains challenging. Direct analysis of Eqs.~\eqref{eq:del_x}--\eqref{eq:del_th} is computationally expensive, especially when one is interested in collective-state statistics or extensive parameter scans. Standard first-order ``delay-to-phase-lag'' reductions are also of limited use here because they miss genuinely high-dimensional effects induced by time delay. By contrast, the second-order framework developed in the main text converts delay into effective inertia and triadic interactions. In this section, we adapt that approach to the time-delayed swarmalator model \eqref{eq:del_x}--\eqref{eq:del_th}.

We consider Eqs.~\eqref{eq:del_x}--\eqref{eq:del_th} in the regime of nonzero mean frequencies with weak heterogeneity around their averages. Specifically, we write
\begin{equation}
	\nu_{k} = \nu + \varepsilon \tilde{\nu}_{k}, \qquad \omega_{k} = \omega + \varepsilon \tilde{\omega}_{k},
\end{equation}
and assume weak coupling
\begin{equation}
	J=\varepsilon \mathcal{J}, \qquad K=\varepsilon\mathcal{K},
\end{equation}
where $\varepsilon\ll 1$ is the small parameter of the perturbation expansion.

To decouple the delayed interaction terms as much as possible, we introduce the sum and difference variables
\begin{equation}
	\phi_{k}(t) = x_{k}(t) + \theta_{k}(t), \qquad \psi_{k}(t) = x_{k}(t) - \theta_{k}(t).
\end{equation}
In these variables, the model takes the form of a pair of linearly coupled delayed Kuramoto-type equations:
\begin{gather}
	\frac{d\phi_{k}}{dt} =
	\Omega_{+} + \varepsilon(\tilde{\nu}_{k}+\tilde{\omega}_{k})
	+ \frac{\varepsilon \mathcal{A}}{N}\sum_{j=1}^{N}\sin\bigl(\phi_{j}(t-\tau)-\phi_{k}-\gamma_{+}\bigr)
	+ \frac{\varepsilon \mathcal{B}}{N}\sum_{j=1}^{N}\sin\bigl(\psi_{j}(t-\tau)-\psi_{k}-\gamma_{-}\bigr), \label{eq:xi_del}\\
	\frac{d\psi_{k}}{dt} =
	\Omega_{-} + \varepsilon(\tilde{\nu}_{k}-\tilde{\omega}_{k})
	+ \frac{\varepsilon \mathcal{B}}{N}\sum_{j=1}^{N}\sin\bigl(\phi_{j}(t-\tau)-\phi_{k}-\gamma_{+}\bigr)
	+ \frac{\varepsilon \mathcal{A}}{N}\sum_{j=1}^{N}\sin\bigl(\psi_{j}(t-\tau)-\psi_{k}-\gamma_{-}\bigr), \label{eq:eta_del}
\end{gather}
where
\begin{equation}
	\mathcal{A}=\frac{\mathcal{J}+\mathcal{K}}{2},\qquad
	\mathcal{B}=\frac{\mathcal{J}-\mathcal{K}}{2},
\end{equation}
\begin{equation}
	\Omega_{+}=\nu+\omega,\qquad \Omega_{-}=\nu-\omega,
\end{equation}
and
\begin{equation}
	\gamma_{+}=\beta+\alpha,\qquad \gamma_{-}=\beta-\alpha.
\end{equation}

To describe macroscopic collective states, we introduce the complex order parameters for the sum and difference variables:
\begin{equation}
	W_+ = \frac{1}{N}\sum_{j=1}^{N} e^{i\phi_j} = r_1^{+} e^{i\vartheta_1^+}, \qquad
	W_- = \frac{1}{N}\sum_{j=1}^{N} e^{i\psi_j} = r_1^{-} e^{i\vartheta_1^-}. \label{eq:orderparams}
\end{equation}
Here, $r_1^{\pm}$ measures the degree of coherence in the two subsystems, while $\vartheta_1^{\pm}$ denotes the corresponding mean phases.

\subsection{Multiple-time-scale expansion and second-order reduction}

We now apply the same multiple-time-scale procedure used for the delayed Kuramoto--Daido network. Introducing slow times $t_s=\varepsilon^s t$, we seek asymptotic solutions of the form
\begin{align}
	\phi_{k}(t) &= \Omega_{+} t_0 + \Phi_{k}(t_1,t_2,\ldots) + \varepsilon \phi_{k}^{(1)}(t_0,t_1,\ldots) + \mathcal{O}(\varepsilon^2), \\
	\psi_{k}(t) &= \Omega_{-} t_0 + \Psi_{k}(t_1,t_2,\ldots) + \varepsilon \psi_{k}^{(1)}(t_0,t_1,\ldots) + \mathcal{O}(\varepsilon^2).
\end{align}
The slow variables $\Phi_k$ and $\Psi_k$ are, respectively, the sum and difference combinations of the slow spatial coordinates $X_k$ and slow oscillator phases $\Theta_k$:
\begin{equation}
	\Phi_{k}(t_1,t_2,\ldots)=X_{k}(t_1,t_2,\ldots)+\Theta_{k}(t_1,t_2,\ldots), \qquad
	\Psi_{k}(t_1,t_2,\ldots)=X_{k}(t_1,t_2,\ldots)-\Theta_{k}(t_1,t_2,\ldots).
\end{equation}

Eliminating secular terms yields the first-order slow dynamics, written in the original time variable $t$, as
\begin{gather}
	\frac{d\Phi_{k}}{dt} = \delta_{k}
	+ \frac{A}{N}\sum_{j=1}^{N}\sin\bigl(\Delta_{jk}^{\phi}\bigr)
	+ \frac{B}{N}\sum_{j=1}^{N}\sin\bigl(\Delta_{jk}^{\psi}\bigr), \label{eq:first_Phi}\\
	\frac{d\Psi_{k}}{dt} = \eta_{k}
	+ \frac{B}{N}\sum_{j=1}^{N}\sin\bigl(\Delta_{jk}^{\phi}\bigr)
	+ \frac{A}{N}\sum_{j=1}^{N}\sin\bigl(\Delta_{jk}^{\psi}\bigr), \label{eq:first_Psi}
\end{gather}
where
\begin{equation}
	A=\varepsilon\mathcal{A}, \qquad B=\varepsilon\mathcal{B},
\end{equation}
\begin{equation}
	\delta_{k}=\varepsilon\left(\tilde{\nu}_{k}+\tilde{\omega}_{k}\right), \qquad
	\eta_{k}=\varepsilon\left(\tilde{\nu}_{k}-\tilde{\omega}_{k}\right),
\end{equation}
and
\begin{gather}
	\Delta_{jk}^\phi(t_1,t_2,\ldots) = \Phi_{j}(t_1,t_2,\ldots)-\Phi_{k}(t_1,t_2,\ldots)-\Omega_{+}\tau-\gamma_{+}, \\
	\Delta_{jk}^\psi(t_1,t_2,\ldots) = \Psi_{j}(t_1,t_2,\ldots)-\Psi_{k}(t_1,t_2,\ldots)-\Omega_{-}\tau-\gamma_{-}.
\end{gather}
Thus, at first order, the delay enters only through the effective phase shifts $\Omega_{+}\tau+\gamma_{+}$ and $\Omega_{-}\tau+\gamma_{-}$.

At second order, however, the delay couples the slow derivatives and generates both inertia and triadic corrections. In the original time variable, the reduced equations take the compact form
\begin{equation}
	\begin{aligned}
		\tau\frac{d^{2}\Phi_{k}}{dt^{2}} + \frac{d\Phi_{k}}{dt} =
		&\left(\delta_{k}
		+ \frac{A}{N}\sum_{j=1}^{N}\sin\bigl(\Delta_{jk}^{\phi}\bigr)
		+ \frac{B}{N}\sum_{j=1}^{N}\sin\bigl(\Delta_{jk}^{\psi}\bigr)\right)
		\left(1 - \frac{\tau A}{N}\sum_{j=1}^{N}\cos\bigl(\Delta_{jk}^{\phi}\bigr)\right) \\
		&\qquad-\left(\frac{\tau B}{N}\sum_{j=1}^{N}\cos\bigl(\Delta_{jk}^{\psi}\bigr)\right)
		\left(\eta_{k}
		+ \frac{B}{N}\sum_{j=1}^{N}\sin\bigl(\Delta_{jk}^{\phi}\bigr)
		+ \frac{A}{N}\sum_{j=1}^{N}\sin\bigl(\Delta_{jk}^{\psi}\bigr)\right),
	\end{aligned}
	\label{eq:second_Phi}
\end{equation}
\begin{equation}
	\begin{aligned}
		\tau\frac{d^{2}\Psi_{k}}{dt^{2}} + \frac{d\Psi_{k}}{dt} =
		&\left(\eta_{k}
		+ \frac{B}{N}\sum_{j=1}^{N}\sin\bigl(\Delta_{jk}^{\phi}\bigr)
		+ \frac{A}{N}\sum_{j=1}^{N}\sin\bigl(\Delta_{jk}^{\psi}\bigr)\right)
		\left(1 - \frac{\tau A}{N}\sum_{j=1}^{N}\cos\bigl(\Delta_{jk}^{\psi}\bigr)\right) \\
		&\qquad-\left(\frac{\tau B}{N}\sum_{j=1}^{N}\cos\bigl(\Delta_{jk}^{\phi}\bigr)\right)
		\left(\delta_{k}
		+ \frac{A}{N}\sum_{j=1}^{N}\sin\bigl(\Delta_{jk}^{\phi}\bigr)
		+ \frac{B}{N}\sum_{j=1}^{N}\sin\bigl(\Delta_{jk}^{\psi}\bigr)\right).
	\end{aligned}
	\label{eq:second_Psi}
\end{equation}

Equations \eqref{eq:second_Phi}--\eqref{eq:second_Psi} are the swarmalator analog of the second-order reduction derived in the main text for time-delayed Kuramoto--Daido networks. The memory encoded by the delay appears as an effective inertial term, while products of mean-field contributions generate higher-order multibody corrections.

\subsection{Back-transformation to slow spatial and phase dynamics}

To recover the reduced dynamics in terms of the slow spatial coordinates $X_k(t)$ and slow oscillator phases $\Theta_k(t)$, we introduce the shifted differences
\begin{gather}
	\Delta_{jk}^{x}(t_1,t_2,\ldots)=X_{j}(t_1,t_2,\ldots)-X_{k}(t_1,t_2,\ldots)-\nu\tau-\beta, \\
	\Delta_{jk}^{\theta}(t_1,t_2,\ldots)=\Theta_{j}(t_1,t_2,\ldots)-\Theta_{k}(t_1,t_2,\ldots)-\omega\tau-\alpha,
\end{gather}
and use trigonometric identities to back-transform Eqs.~\eqref{eq:second_Phi}--\eqref{eq:second_Psi}. This yields
\begin{align}
	\tau\frac{d^{2}X_{k}}{dt^{2}} + \frac{dX_{k}}{dt} =&
	\left(1-\frac{\tau J}{N}\sum_{j=1}^{N} \cos\bigl(\Delta_{jk}^{x}\bigr)\cos\bigl(\Delta_{jk}^{\theta}\bigr)\right)\!
	\left(\varepsilon\tilde\nu_{k} + \frac{J}{N}\sum_{j=1}^{N} \sin\bigl(\Delta_{jk}^{x}\bigr)\cos\bigl(\Delta_{jk}^{\theta}\bigr)\right) \nonumber \\
	&+ \left(\frac{\tau J}{N}\sum_{j=1}^{N} \sin\bigl(\Delta_{jk}^{x}\bigr)\sin\bigl(\Delta_{jk}^{\theta}\bigr)\right)\!
	\left(\varepsilon\tilde\omega_{k} + \frac{K}{N}\sum_{j=1}^{N} \cos\bigl(\Delta_{jk}^{x}\bigr)\sin\bigl(\Delta_{jk}^{\theta}\bigr)\right), \label{eq:ddotX}\\
	\tau\frac{d^{2}\Theta_{k}}{dt^{2}} + \frac{d\Theta_{k}}{dt} =&
	\left(1 - \frac{\tau K}{N}\sum_{j=1}^{N} \cos\bigl(\Delta_{jk}^{x}\bigr)\cos\bigl(\Delta_{jk}^{\theta}\bigr)\right)\!
	\left(\varepsilon\tilde{\omega}_{k} + \frac{K}{N}\sum_{j=1}^{N} \cos\bigl(\Delta_{jk}^{x}\bigr)\sin\bigl(\Delta_{jk}^{\theta}\bigr)\right) \nonumber \\
	&+ \left(\frac{\tau K}{N}\sum_{j=1}^{N} \sin\bigl(\Delta_{jk}^{x}\bigr)\sin\bigl(\Delta_{jk}^{\theta}\bigr)\right)\!
	\left(\varepsilon\tilde{\nu}_{k} + \frac{J}{N}\sum_{j=1}^{N} \sin\bigl(\Delta_{jk}^{x}\bigr)\cos\bigl(\Delta_{jk}^{\theta}\bigr)\right). \label{eq:ddotTh}
\end{align}

Thus, although the original swarmalator model contains only delayed pairwise coupling, the second-order delay-free reduction acquires effective hypernetwork-like terms.

For analytical convenience and numerical implementation, it is useful to introduce the ensemble-averaged combinations
\begin{align}
	S^{cc}_k &= \frac{1}{N}\sum_{j=1}^{N}\cos\bigl(\Delta_{jk}^x\bigr)\cos\bigl(\Delta_{jk}^\theta\bigr), \qquad
	S^{sc}_k = \frac{1}{N}\sum_{j=1}^{N}\sin\bigl(\Delta_{jk}^x\bigr)\cos\bigl(\Delta_{jk}^\theta\bigr), \notag\\[4pt]
	S^{ss}_k &= \frac{1}{N}\sum_{j=1}^{N}\sin\bigl(\Delta_{jk}^x\bigr)\sin\bigl(\Delta_{jk}^\theta\bigr), \qquad
	S^{cs}_k = \frac{1}{N}\sum_{j=1}^{N}\cos\bigl(\Delta_{jk}^x\bigr)\sin\bigl(\Delta_{jk}^\theta\bigr). \notag
\end{align}
In terms of these quantities, Eqs.~\eqref{eq:ddotX} and \eqref{eq:ddotTh} become
\begin{align}
	\tau \frac{d^2 X_k}{dt^2} + \frac{dX_k}{dt} &= \bigl(1 - \tau J S^{cc}_k\bigr)\bigl(\varepsilon\tilde{\nu}_k + J S^{sc}_k\bigr) + \tau J S^{ss}_k \bigl(\varepsilon\tilde{\omega}_k + K S^{cs}_k\bigr), \label{eq:compactX}\\[4pt]
	\tau \frac{d^2 \Theta_k}{dt^2} + \frac{d\Theta_k}{dt} &= \bigl(1 - \tau K S^{cc}_k\bigr)\bigl(\varepsilon\tilde{\omega}_k + K S^{cs}_k\bigr) + \tau K S^{ss}_k \bigl(\varepsilon\tilde{\nu}_k + J S^{sc}_k\bigr). \label{eq:compactTheta}
\end{align}

In this form, the role of the delay is especially transparent. The factors $(1-\tau J S^{cc}_k)$ and $(1-\tau K S^{cc}_k)$ describe a delay-induced renormalization of the response to the collective field. The cross terms proportional to $\tau S^{ss}_k$ are genuine second-order contributions and represent triadic couplings between the spatial and phase subsystems. The quantities $S^{cc}_k$, $S^{sc}_k$, $S^{ss}_k$, and $S^{cs}_k$ can in turn be expressed through the real and imaginary parts of $W_{\pm}$ and their spatial modulations, linking the delay-induced corrections in Eqs.~\eqref{eq:compactX}--\eqref{eq:compactTheta} directly to the global order parameters.

\end{document}